\documentclass[oneside,12pt,margin=1in]{article}
\usepackage[utf8]{inputenc}
\usepackage{rotating}
\usepackage[left = 1.2in, top = 1in, right = 1.2in]{geometry} 
\usepackage{setspace}
\usepackage{natbib}
\bibpunct{(}{)}{;}{a}{,}{,}
\usepackage{amsmath}
\usepackage{amssymb}
\usepackage{amsthm}
\usepackage{bm}
\usepackage[usenames,dvipsnames]{xcolor}
\usepackage{multirow}
\usepackage{subfig}
\usepackage{tikz}
\usepackage{refcheck}
\usepackage{colortbl}
\usepackage{longtable}
\usepackage{placeins}
\usepackage{subfig}

\usetikzlibrary{arrows,shapes,positioning,petri}
\def\m{{\bf m}}

\def\t{{\bf t}}

\def\w{{\bf w}}
\def\X{{\bf X}}
\def\x{{\bf x}}

\def\z{{\bf z}}

\def\0{{\bf 0}}
\def\1{{\bf 1}}

\def\<{\, \langle \,}
\def\>{\, \rangle \,}
\def\inv{^{-1}}
\def\ts{^\top}

\def\bPsi{\bm{\Psi}}
\def\bpi{\bm{\pi}}

\def\bepsilon{\bm{\epsilon}}

\def\btheta{\bm{\theta}}

\def\bPhi{\bm{\Phi}}

\def\DN{\mathcal{N}}

\def\GIG{\mathrm{GIG}}
\def\tGIG{\mathrm{tGIG}}

\def\Exp{\mathrm{Exponential}}

\def\Dis{\mathrm{Discrete}}

\newcommand{\argmin}{\operatornamewithlimits{argmin}}

\newcommand{\ie}{i.e.\ }

\newcommand{\iid}{i.i.d.\ }

\def\ctilde{\kern -.04em\lower .7ex\hbox{\~{}}\kern .04em}

\newtheorem{lemma}{Lemma}
\begin{document}
\title{Efficient model-based clustering with coalescents: Application to multiple outcomes using medical records data
\footnotetext[0]{Ricardo Henao and Joseph E. Lucas are Assistant Research Professors in Electrical and Computer Engineering, Duke University, Durham, NC 27710. E-mail: r.henao@duke.edu and joe@stat.duke.edu.}
}
\author{Ricardo Henao and Joseph E. Lucas}
%
\date{\small \today}
\maketitle
\doublespacing
\begin{abstract}
	We present a sequential Monte Carlo sampler for coalescent based Bayesian hierarchical clustering.  The model is appropriate for multivariate non-\iid data and our approach offers a substantial reduction in computational cost when compared to the original sampler. We also propose a quadratic complexity approximation that in practice shows almost no loss in performance compared to its counterpart. Our formulation leads to a greedy algorithm that exhibits performance improvement over other greedy algorithms, particularly in small data sets. We incorporate the Coalescent into a hierarchical regression model that allows joint modeling of multiple correlated outcomes. The approach does not require {\em a priori} knowledge of either the degree or structure of the correlation and, as a byproduct, generates additional models for a subset of the composite outcomes.  We demonstrate the utility of the approach by predicting multiple different types of outcomes using medical records data from a cohort of diabetic patients.
	
\end{abstract}
\section{Introduction}
Learning hierarchical structures from observed data is a common practice in many knowledge domains. Examples include phylogenies and signaling pathways in biology, language models in linguistics, etc.  Agglomerative clustering is still the most popular approach to hierarchical clustering due to its efficiency, ease of implementation and a wide range of possible distance metrics.  However, because it is algorithmic in nature, there is no principled way that agglomerative clustering can be used as a building block in more complex models. Bayesian priors for structure learning on the other hand, are perfectly suited to be employed in larger models. Several authors have proposed using hierarchical structure priors to model correlation in factor models \citep{rai08,zhang11a,henao12a,henao13a}.

There are many approaches to hierarchical structure learning already proposed in the literature, see for instance \citet{neal03,heller05,teh08,adams10}. The work in this paper focuses on the Bayesian agglomerative clustering model proposed by \citet{teh08}. This allows us to perform model based hierarchical clustering with continuous multivariate non \iid data -- by which we mean multivariate observations in which the elements of the vector are not \iid. Although the authors introduce priors both for continuous and discrete data, no attention is paid to the non \iid case, mainly because their work is focused in proposing different inference alternatives.

Kingman's coalescent is a standard model from population genetics perfectly suited for hierarchical clustering since it defines a prior over binary trees \citep{kingman82,kingman82a}. This work advances Bayesian hierarchical clustering in two ways: (i) we extend the original model to handle non \iid data and (ii) we propose an efficient sequential Monte Carlo inference procedure for the model which scales quadratically rather than cubically. As a byproduct of our approach we propose as well a small correction to the greedy algorithm of \citet{teh08} that shows gains particularly in small data sets.  In addition, we demonstrate a novel application of the Coalescent within a hierarchical regression that gives improved predictive accuracy in the presence of multiple correlated outcomes.

There is a separate approach by \citet{gorur08} that also improves the cubic computational cost of Bayesian hierarchical clustering. They introduce an efficient sampler with quadratic cost that although proposed for discrete data can be easily extended to continuous data, however as we will show, our approach is still considerably faster.

The remainder of the manuscript is organized as follows, the data model and the use of coalescents as priors for hierarchical clustering are reviewed in Section~\ref{sc:coalescent}. Our approach to inference and relationships to previous approaches are described in Section~\ref{sc:inference}. Section \ref{sec:emrResults} demonstrates the use of the model and algorithm in the context of hierarchical regression with multiple outcomes of unknown correlation.  The appendix contains numerical results on both artificial and real data.
\section{Coalescents for hierarchical clustering} \label{sc:coalescent}
A {\em partition} of a set $A$ is defined to be a collection of subsets $\{a_j\}_{j=1}^J$ such that (1) $a_j\subset A$ for all $j$, (2) $\cup_j a_j = A$ and (3) $a_j \cap a_{j'} = \varnothing$ for all $j\neq j'$.  Let $X$ be an $n\times d$ dimensional matrix of $n$ observations in $d$ dimensions with rows $\x_i$.  We will define a binary tree on our set of $n$ observations; define $\pi_0 = \{\{\x_1\}, \ldots, \{\x_n\}\}$ to be a partition containing $n$ singletons. We iteratively combine pairs of elements such that $\pi_{i+1}$ is obtained by removing two of the elements from $\pi_i$ and inserting their union.  Thus if $\pi_i = \{a_j\}_{j=1}^{n-i}$ then $\pi_{i+1}=(\pi_i \setminus \{a_{j},a_{j'}\}) \cup (a_{j}\cup a_{j'})$ for some $j\neq j'$.  For $a_j,a_{j'}\in \pi_i$ define $z_{i+1}^*=a_j \cup a_{j'}$ where $j$ and $j'$ indicate the elements of $\pi_i$ that were merged to obtain $\pi_{i+1}$. A particular sequence of $n$ partitions defines a binary tree.  The $n$ leaves of the tree are the elements of $\pi_0$ and the nodes (branching points) of the tree are $\{z_k^*\}_{k=1}^{n-1}$. 

Defining $\t=[t_1 \ \ldots \ t_{n-1}]$ to be a vector of branching times, $\bpi=\{\pi_1,\ldots,\pi_{n-1}\}$ to be the ordered collection of partitions and $\z=\{\z_1, \ldots, \z_{n-1}\}$ to be the latent $d$-dimensional observations at the internal nodes we can write a Bayesian model for hierarchical clustering as
\begin{align} \label{eq:model}
	\begin{aligned}
	\x_i|\t,\bpi \ \sim & \ p(\x_i|\t,\bpi,\z) \,, \\
	\z \ \sim & \ p(\z) \,, \\
	\t,\bpi \ \sim & \ {\rm Coalescent}(n) \,,
	\end{aligned}
\end{align}
$p(\x_i|\t,\bpi,\z)$ is the data likelihood and $p(\z)$ is the prior distribution for the internal nodes of the tree and the pair $\{\t,\bpi\}$ is provided with a prior distribution over binary tree structures known as Kingman's coalescent.

\subsection{Kingman's coalescent}
The $n$-coalescent is a continuous-time Markov chain originally introduced to describe the common genealogy of a sample of $n$ individuals backwards in time \citep{kingman82,kingman82a}. It defines a prior over binary trees with $n$ leaves, one for each individual. The coalescent assumes a uniform prior over tree structures, $\bpi$, and exponential priors on the set of $n-1$ merging times, $\t$.

It can be thought of as a generative process on partitions of $\{1,\ldots,n\}$ as follows
\begin{itemize}
	\item Set $k=1$, $t_0=0$, $\pi_0=\{\{1\},\ldots,\{n\}\}$.
	\item While $k<n$
	\begin{itemize}
		\item Draw $\Delta_k \sim \Exp(\lambda_k) \mbox{ where } \lambda_k=(n-k+1)(n-k)/2$ is a rate parameter.
		\item Set $t_k=t_{k-1} + \Delta_k$.
		\item Merge uniformly two sets of $\pi_{k-1}$ into $\pi_{k}$.
		\item Set $k=k+1$.
	\end{itemize}
\end{itemize}
Because there are $(n-k+1)(n-k)/2$ possible merges at stage $k$, any particular pair in $\pi_i$ merges with prior expected rate 1 for any $i$. We can compute the prior probability of a particular configuration of the pair $\{\t,\bpi\}$ as
\begin{align} \label{eq:prior}
	p(\t,\bpi) = \prod_{k=1}^{n-1} \exp\left(-\tfrac{(n-k+1)(n-k)}{2}\Delta_k\right) \,,
\end{align}
this is, the product of merging and coalescing time probabilities. Some properties of the $n$-coalescent include: (i) $\bpi$ is uniform and independent of $\t$, (ii) it is independent of the order of sets in partition $\pi_i$ for every $i$ and (iii) the expected value of $t_{n-1}$ (last coalescing time) is $\mathbb{ E}[t_{n-1}]=2(1-n\inv)$.
\subsubsection{Distribution of the latent nodes}
Recall that $z_k^*\in \pi_k$ is an internal node in a binary tree with associated latent $d$-dimensional vector $\z_k$.  Let $z^*_{c_1}$ and $z^*_{c_2}$ be the children of $z_k^*$ and define $C=\{c_1, c_2\}$.  We designate the leaves of the tree with $x_i^*$.  If we define $p(\z_k|\z_c,\t)$ to be the {\em transition density} between a child node, $\z_c$, and its parent, $\z_k$, then we can recursively define $q(\z_k|\bpi,\t)$ to be a distribution of $\z_k$ as follows:
\begin{align*}
q(\x_i|\bpi,\t) = & \ \delta_{\x_i} \,, \\
q'(\z_k|\bpi,\t) = & \ \prod_{c\in C}\int p(\z_k|\z_c,\t)q(\z_c|\bpi,\t)d\z_c \,, \\
= & \ Z_k(\X,\bpi, \t) q(\z_k|\bpi,\t) \,,
\end{align*}
where $q(\z_k|\bpi,\t)$ is a density (integrating to 1) and $Z_k$ is the appropriate scaling factor.

Recently, \citet{teh08} showed that by using an agglomerative approach for constructing $\{\t,\bpi\}$, the likelihood for the model in equation~\eqref{eq:model} can be recursively written as
\begin{align} \label{eq:lik}
	p(\X|\t,\bpi) = \prod_{k=1}^{n-1} Z_{k}(\X|\bpi,\t) \,.
\end{align}
Because of the tree structure, $\z_k$ is independent of $\X$ conditional on the distributions of its two child nodes.  This implies that $Z_k(\X|\bpi,\t) = Z_k(\X|\pi_{1:k},\t_{1:k})$.  Our formulation is equivalent to using \emph{message passing} to marginalize recursively from the leaves to the root of the tree \citep{pearl88}. The message is $q(\z_k|\bpi,\t)$ for node $z^*_k$ and it summarizes the entire subtree below node $z^*_k$. 

\begin{figure}[!t]
	\centering
	\begin{tikzpicture}[ bend angle = 5, >=latex, font = \footnotesize ]
		\tikzstyle{lps} = [ circle, thick, draw = black!80, fill = OliveGreen, minimum size = 1mm, inner sep = 2pt ]
		\tikzstyle{ety} = [ minimum size = 1mm, inner sep = 2pt ]
		\begin{scope}[node distance = 1.5cm and 1cm]
			\node [ety] (t_) [ ] {};
			\node [ety] (t0) [ right of = t_, node distance = 2.5cm ] {$t_1=0$};
			\node [ety] (t1) [ right of = t0 ] {$t_2$};
			\node [ety] (t2) [ right of = t1 ] {$t_{k-1}$};
			\node [ety] (t3) [ right of = t2 ] {};
			\node [ety] (t4) [ right of = t3 ] {$t_k$};
			\node [ety] (t5) [ right of = t4 ] {$t_{k+1}$};
			\node [lps] (lp_1) at (2.5,-0.5) [label = -90:$\x_1$] {};
			\node [lps] (lp_2) at (2.5,-1.5) [label = -90:$\x_2$] {};
			\node [ety] (lp_3) at (2.5,-2.5) [label = -90:$\vdots$] {};
			\node [lps] (lp_4) at (2.5,-4.5) [label = -90:$\x_n$] {};
			\node [lps] (v1) at (4.0,-1.0) [ label = -90:$\z_{c_1}$ ] {}
				edge [pre, bend right] (lp_1) 
				edge [pre, bend left] (lp_2);
			\node[ety] (ve1) at (4.0,-3.0) [ label = 180:$\ldots$ ] {};
			\node[ety] (ve2) at (4.0,-4.0) [ label = 180:$\ldots$ ] {};
			\node [lps] (v2) at (5.5,-3.5) [ label = 90:$\z_{c_2}$ ] {}
				edge [pre, bend right] (ve1)
				edge [pre, bend left] (ve2);
			\node [lps] (v4) at (8.5,-2.5) [ label = -90:$\z_k$ ] {}
				edge [pre, bend right] (v1)
				edge [pre, bend left] (v2);
			\node [ety] (v5) at (10.0,-2.5) [ label = 0:$\ldots$ ] {}
				edge [pre] (v4);
			\node [ety] at (4.9,-0.6) [] {$q(\z_{c_1}|c_1,\t_{1:2})$};
			\node [ety] at (6.6,-3.9) [] {$q(\z_{c_2}|c_2,\t_{1:k-1})$};
			\node [ety] at (9.4,-2.1) [] {$q(\z_k|C,\t_{1:k})$};
			\node [ety] at (7.5,-1.3) [] {$p(\z_k|\z_{c_1},\t_{1:k})$};
			\node [ety] at (8.2,-3.35) [] {$p(\z_k|\z_{c_2},\t_{1:k})$};
		\end{scope}
	\end{tikzpicture}
	\caption[Segment of the binary tree structure]{Binary tree structure. Latent variable $\t$ and $\bpi$ define merging points and merging sets, respectively.}
	\label{fg:stree}
\end{figure}
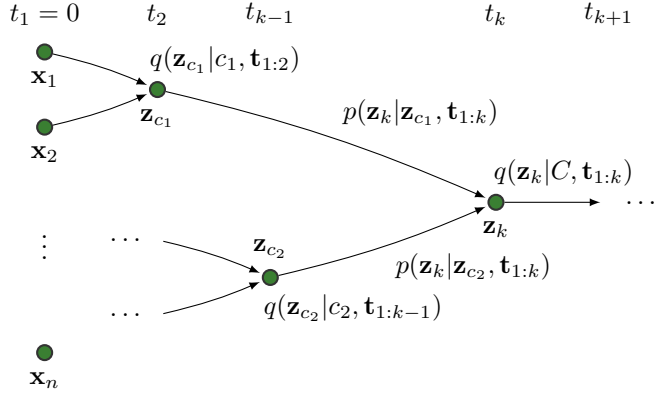

Figure~\ref{fg:stree} illustrates the process for a segment of a tree. The size of partitions $\pi_k$ shrink as $k$ increases, so from the illustration $\pi_0=\{\{1\},\{2\},\ldots,\{n\}\}$, $\pi_1=\{\{1,2\},\ldots,\{n\}\}$, $\pi_{k-1}=\{c_1,c_2,\ldots\}$ and $\pi_k=\{c_1\cup c_2,\ldots\}$.

The joint distribution needed to perform inference can be obtained by combining likelihood and prior in equations \eqref{eq:lik} and \eqref{eq:prior} as
\begin{align} \label{eq:joint}
	p(\X,\t,\bpi) = \prod_{k=1}^{n-1} \exp(-\lambda_k\Delta_k) Z_{k}(\X|\pi_{1:k},\t_{1:k}) \,.
\end{align}
\subsection{Gaussian transition distributions}
A common approach to correlated continuous data is the use of a multivariate Gaussian distribution for the transition density,
\begin{align*} 
	p(\z_k|\z_c,\t_{1:k}) = \DN(\z_k|\z_c,\Delta_{k,c}\bPhi) \,,
\end{align*}
where $\bPhi$ is a covariance matrix encoding the correlation structure in $\z_k$ and $\Delta_{k,c}$ is the time elapsed between $t_k$ and $t_c$, not necessarily $t_{k}-t_{k-1}$ as can be seen in Figure~\ref{fg:stree}. We denote the time at which node $z_c^*$ was created as $t_c$. Individual terms of the likelihood in equation~\eqref{eq:lik} can be computed using
\begin{align*}
	q(\z_k|\pi_{1:k},\t_{1:k}) = & \ \DN(\z_k|\m_{c_1},\tilde{s}_{c_1}\bPhi)\DN(\z_k|\m_{c_2},\tilde{s}_{c_2}\bPhi) \,, \\
	= & \ Z_{k}(\X|\pi_{1:k},\t_{1:k}) \DN(\z_k|s_k(\tilde{s}_{c_1}\inv\m_{c_1}+\tilde{s}_{c_2}\inv\m_{c_2}),s_k\bPhi) \,,
\end{align*}
where $\m_{c_1}$ and $s_{c_1}$ are mean and variance of $q(\z_{c_1}|\bpi,\t)$, respectively. Furthermore, $\tilde{s}_{c_1}=\Delta_{k,c_1}+s_{c_1}$.  This leads to $s_k=(\tilde{s}_{c_1}\inv+\tilde{s}_{c_2}\inv)\inv$ and the normalization constant is
\begin{align}
	Z_{k}(\X|\pi_{1:k},\t_{1:k}) = & (2\pi)^{-d/2}|v_k\bPhi|^{-1/2}\exp\left(-\tfrac{1}{2}(\m_{c_1}-\m_{c_2})v_k\inv\bPhi\inv(\m_{c_1}-\m_{c_2})\ts\right)  \,, \label{eq:ZkNorm}
\end{align}
where $v_k=\tilde{s}_{c_1}+\tilde{s}_{c_2}=2\Delta_k+r_k$ and $r_k=2t_{k-1}-t_{c_1}-t_{c_2}+s_{c_1}+s_{c_2}$. Note that $\Delta_{c_1}=\Delta_{c_2}$ only if $c_1$ and $c_2$ are singletons. The term $r_k$ can be interpreted as the accumulated variance up to $t_{k-1}$ given partition $\pi_{k-1}$, \ie it summarizes the two subtrees encompassed by sets $c_1$ and $c_2$.
\section{Inference} \label{sc:inference}
Inference is carried out using a sequential Monte Carlo (SMC) sampling based upon equation~\eqref{eq:joint} \citep[see][]{doucet01}. Define $\mathcal{C}_i$ to be the set of all pairs of elements from the partition $\pi_i$, and for $C\in \mathcal{C}_i$ define $\pi_{i,C}$ to be the partition obtained by merging the two elements in $C$.  For a set of $M$ particles, we approximate the posterior of the pair $\{\t,\bpi\}$ using a weighted sum of point masses obtained iteratively by drawing coalescing times $t_k$ and chain states $\pi_k$ one at a time from their posterior as
\begin{align}
	p(\Delta_k,\pi_{k-1,C}|\t_{1:k-1},\pi_{1:k-1}) \ = & \ Z\inv\exp(-\lambda_k\Delta_k) Z_{k}(\X|\pi_{1:k-1},\pi_{k-1,C},\t_{1:k-1},\Delta_k) \,, \label{eq:delta_pi_post} \\
	Z \ = & \sum_{c \in\mathcal{C}_k} \underbrace{\int \exp(-\lambda_k\Delta_k)Z_k(\X|\pi_{1:k-1},\pi_{k-1,c},\t_{1:k-1},\Delta_k) d\Delta_k}_{Z_{k,c}} \,, \label{eq:Zkc}
\end{align}
%

From equation~\eqref{eq:Zkc} we see that the integral needs to be computed for every pair in $\pi_{k-1}$ at every iteration of the sampler, simply because the rate of the exponential distribution, $\lambda_k$, is a function of $k$. Algorithms introduced by \citet{teh08} try to avoid the computational complexity of using equation \eqref{eq:delta_pi_post} directly by simplifying it or by means of greedy alternatives. They propose for instance to draw $\Delta_k$ from the prior; then computing $Z_{k,C}$ is no longer necessary thus reducing computational cost. We will show that by using some properties of the distributions involved in equation~\eqref{eq:delta_pi_post} we can effectively decrease the computational complexity of the SMC sampler with almost no performance penalty. In particular, we will show that the most expensive parts of equations~\eqref{eq:delta_pi_post} and \eqref{eq:Zkc} need to be computed only once during inference.

If we assume a Gaussian transition distribution, we can rewrite and then expand equation~\eqref{eq:delta_pi_post} as
\begin{align}
	 p(\Delta_k,\pi_{k-1,C}|\t_{1:k-1},\pi_{1:k-1}) \ \propto & \ Z_{k}(\X|\pi_{k-1},C,\t_{1:k})\exp\left(-\frac{\lambda_k}{2}(2\Delta_k+r_k)\right)\exp\left(\frac{\lambda_k}{2}r_k\right) \,, \nonumber \\
	\propto & \ \tGIG_{r_k}(2\Delta_k+r_k|\bepsilon_{k-1,C},\lambda_k)Z_{k,C} \,, \label{eq:delta_pi_modpost}
\end{align}
where $\bepsilon_{k-1,C}=(\m_{c_1}-\m_{c_2})\bPhi\inv(\m_{c_1}-\m_{c_2})\ts$ and $\tGIG_{r_k}(\lambda_k,\chi,\psi)$ is the generalized inverse Gaussian (GIG) with parameters $\{\lambda_k,\chi,\psi\}$ truncated below $r_k$ \citep{jorgensen82a} and the last term to the right hand side of equation~\eqref{eq:delta_pi_modpost} is used to denote that pair $C$ is selected with probability proportional to $Z_{k,C}$.
\begin{align} \label{eq:pi_modpost}
	Z_{k,C} \ \approx & \ \frac{K_{1-d/2}(\sqrt{\lambda_k\bepsilon_{k-1,C}})}{(\lambda_k\bepsilon_{k-1,C}\inv)^{(1-d/2)/2}}\exp\left(\frac{\lambda_k}{2} r_k\right) \,,
\end{align}
where $K_{\nu}(z)$, the modified Bessel function of second kind \citep{abramowitz65a}, is the normalizer of the generalized inverse Gaussian distribution. Since the normalization constant of the truncated GIG distribution does not have a closed form and assuming that the posterior distribution of $\Delta_k$ is peaked, we approximate the true normalizer with $K_{\nu}(z)$. Later in the paper we explore empirically the effects of this approximation with some artificially generated data. The details on how to obtain equations~\eqref{eq:delta_pi_modpost} and \eqref{eq:pi_modpost} can be found in Appendix~\ref{eq:delta_pi_details}. 

From Equation~\eqref{eq:delta_pi_modpost} we have
\begin{align}
	\Delta_k|C,\t_{1:k-1} \ \sim & \ \tGIG_{r_k}(2\Delta_k+r_k|\bepsilon_{k,C},\lambda_k) \,, \label{eq:delta_sample} \\
	\pi_k|\pi_{k-1},\t_{1:k-1} \ \sim & \ \Dis(\pi_{k-1,C}|\w_{k-1}) \,, \label{eq:pi_sample}
\end{align}
where $\w_{k-1}$ is the vector of normalized weights ranging over $\mathcal{C}_{k-1}$, computed using equation~\eqref{eq:pi_modpost}. Sampling from equations~\eqref{eq:pi_modpost}, \eqref{eq:delta_sample} and \eqref{eq:pi_sample} have useful properties, (i) the conditional posterior of $\pi_k$ does not depend on $\Delta_k$. (ii) Sampling from $\Delta_k$ amounts to drawing from a truncated generalized inverse Gaussian distribution. (iii) We do not need to sample $\Delta_k$ for every pair in $\pi_{k-1}$, in fact we only need to do so for the merging pair. (iv) Although $\lambda_k$ in equation~\eqref{eq:pi_modpost} changes with $k$, the most expensive computation, $\bepsilon_{k-1,C}$, needs to be computed only once. (v) The expression in equation~\eqref{eq:pi_modpost} can be seen as the core of a distribution for $\m_{c_1}-\m_{c_2}$ which has heavier tails than a Gaussian distribution, and $Z_{k,C}\to\infty$ as $\bepsilon_{k-1,C}\to0$ for $d>1$. Furthermore, we can rewrite equation~\eqref{eq:pi_modpost} as
\begin{align}
	Z_{k,C} \ \propto & \ \frac{K_{d/2-1}(\sqrt{\lambda_k\bepsilon_{k-1,C}})}{(\lambda_k\inv\bepsilon_{k-1,C})^{(d-2)/4}}\exp\left(\frac{\lambda_k}{2} r_k\right) \,, \label{eq:pi_modpost_rep} \\
	Z_{k,C} \ \approx & \ \bepsilon_{k-1,C}^{-(d-1)/4}\exp(-\sqrt{\lambda_k\bepsilon_{k-1,C}})\exp\left(\frac{\lambda_k}{2} r_k\right) \label{eq:pi_modpost_ne} \,,
\end{align}
where in equation~\eqref{eq:pi_modpost_rep} we have made a change of variables before marginalizing out $v_k=2\Delta_k+r_k$, and in equation~\eqref{eq:pi_modpost_ne} we have used the limiting form of $K_\nu(z)$ as $z\to\infty$ \citep{abramowitz65a}. \citet{eltoft06a} have called equation~\eqref{eq:pi_modpost_rep} the core of their multivariate Laplace distribution, $\m_{c_1}-\m_{c_2}\sim{\rm ML}(\lambda_k,\bPsi)$ with parameters $\lambda_k$ and $\bPsi$. When $d=1$ equation~\eqref{eq:pi_modpost_ne} is no longer an approximation; it is the core of a univariate Laplace distribution with rate $\sqrt{\lambda_k\bPsi\inv}$. Additionally, equation~\eqref{eq:pi_modpost_ne} can be particularly useful as a cheap numerically stable alternative to $K_\nu(z)$ when $d$ is large.
\subsection{Sampling coalescing times}
Sampling from a generalized inverse Gaussian distribution is traditionally done using the \emph{ratio-of-uniforms} method of \citet{dagpunar89a}. We observed empirically that a slice sampler within the interval $(r_k/2,\Delta_0r_k/2)$ is considerably faster than the commonly used algorithm. Although we use $\Delta_0=10^2$ in all our experiments, we did try larger values without noticing significant changes in the results. The slice sampler used here is a standard implementation of the algorithm described by \citet{neal03a}. We acknowledge that adaptively selecting $\Delta_0$ at each step could improve the efficiency of the sampler however we did not investigate it.
\subsection{Covariance matrix}
Until now we assumed the covariance matrix $\bPhi$ as known, in most cases however the correlation structure of the observed data is unavailable. For unknown $\bPhi$ we alternate between SMC sampling for the tree structure and drawing $\bPhi$ from some suitable distribution.  We do this by repeating the procedure for a number of iterations ($N_{\rm iter}$) and then dropping a subset of these as burn-in period. Because averaging of tree structures is not a well defined operation, after summarizing the posterior samples of the hyperparameters of the covariance function using medians, we perform a final SMC step to obtain a final tree structure. 

For cases when observations exhibit additional structure, such as temporal or spatial data, we may assume the latent variable $\z_k$ is drawn from a Gaussian process.  We suppose that elements of $\bPhi$ are computed using $\phi_{ij}=g(i,j,\btheta)$ for a set of hyperparameters $\btheta$. For example, we could use a squared exponential covariance function
\begin{align} \label{eq:sqexp_noise}
	g(i,j,\ell,\sigma^2) = \ \exp\left(-\frac{1}{2\ell}d_{ij}^2\right) + \sigma^2\delta_{ij} \,,
\end{align}
where $\btheta=\{\ell,\sigma^2\}$, $\delta_{ij}=1$ IFF $i=j$ and $d_{ij}$ is the time between samples $i$ and $j$. The smoothness of the process is controlled by the inverse length scale $\ell$ and the amount of idiosyncratic noise by $\sigma^2$. The elements of $\btheta$ are sampled by coordinate-wise slice sampling using the following function as proxy for the elements of $\btheta$,
\begin{align*}
	f(\btheta|\pi,\t) = \ \sum_{k=1}^{n-1} Z_{k}(\X|\pi_{1:k},\t_{1:k}) \,.
\end{align*}

This approach is generally appropriate for continuous signals where smoothing of the covariance estimates is desirable.  For the case when no smoothness is required but correlation structure is expected, conjugate inverse Wishart distributions for $\bPhi$ can be considered. For \iid data, a diagonal/spherical $\bPhi$ with independent inverse gamma priors is a good choice, as already proposed by \citet{teh08}.

\subsection{Greedy implementation}
Here we propose a method to draw a single sample with high posterior likelihood from the Coalescent. Such a sample can be built by greedily maximizing equation~\eqref{eq:joint} one step at the time. This requires the computation of the mode of $\Delta_k$ from equation~\eqref{eq:delta_sample} for every pair in $\pi_{k-1}$ and merging the pair with smallest $\Delta_k$.  For a given pair, $C$ we have
$$\mbox{mode}(\Delta_{k}|C) = \ \frac{1}{2\lambda_k}\left(-d/2+\sqrt{d^2/4+\lambda_k\bepsilon_{k-1,C}}\right)-\frac{1}{2}r_k $$
and the algorithm selects $C$ such that
$$C = \argmin_{C'}\{\Delta_{k,C'},C'\in\mathcal{C}_{k-1}\}.$$
Because the posterior of $\Delta_k$ is skewed to the right, a greedy implementation based on the mode of the distribution will on average underestimate coalescing times. If we use the posterior mean this bias can be decreased. We then propose using
\begin{align*}
	\mbox{mean}(\Delta_{k}|C) = \ \frac{1}{2}\underbrace{\sqrt{\frac{\bepsilon_{k-1,C}}{\lambda_k}}\frac{K_{2-d/2}(\sqrt{\lambda_k\bepsilon_{k-1,C}})}{K_{1-d/2}(\sqrt{\lambda_k\bepsilon_{k-1,C}})}}_{\mu_{\Delta_k}} -\frac{1}{2}r_k \,,
\end{align*}
where $\mu_{\Delta_k}$ is the mean of $\GIG(2\Delta_k+r_k|\bepsilon_{k,C},\lambda_k)$. We use $\mu_{\Delta_k}$ as an approximation to the mean of the truncated generalized inverse Gaussian distribution because no closed form is available. This means that although our approximation will be also biased to the left of the true coalescing time, it will be considerably less so than the proposal based on the mode of the distribution proposed by \citet{teh08}.

We expect significant differences between the two greedy approaches only for low dimensional data sets because as $d$ becomes large, the posterior of $\Delta_k$ will be highly peaked thus making the distance between mean and mode too small to make the proposals distinguishable.

Computing two modified Bessel functions by brute force is more expensive and numerically unstable compared to the simple proposal based on the mode of the GIG distribution, however we can recursively compute the ratio of Bessel functions using the identity $K_{v+1}(z) = K_{v-1} + 2vz\inv K_{v}(z)$, \citep{abramowitz65a} thus
\begin{align*}
	\frac{K_{v+1}(z)}{K_v} = \frac{K_{v-1}(z)}{K_v} + \frac{2v}{z} \,,
\end{align*}
starting from the closed forms of $K_{-0.5}$ and $K_{0.5}$ if $d$ is even, or rational approximations (accurate to 19 digits \citep{blair74a}) to $K_{1}$ and $K_{0}$ if $d$ is odd. 

Incidentally, a similar recursion can be used to compute the variance of the GIG distribution at almost no cost since it is also a function of ratios involving $K_{v+2}(z)$, $K_{v+1}(z)$ and $K_{v}(z)$.  The combination of mean and variance estimates can be used to estimate the mass of the distribution that is lost by truncating at $r_k$.

\subsection{Computational cost}
The computational cost of using equation \eqref{eq:delta_pi_post} directly to sample from $\t$ and $\bpi$ for a single particle is ${\cal O}(\kappa_1n^3)$, where $\kappa_1$ is the cost of drawing the merging time of a single candidate pair \citep{teh08}. Using equation \eqref{eq:delta_pi_modpost} costs ${\cal O}(\kappa_2n^3 + \kappa_1n)$, where $\kappa_2$ is the cost of computing $Z_{k,C}$ for a single candidate pair. Since $\kappa_2<<\kappa_1$, using equation \eqref{eq:delta_pi_modpost} is much faster than previous approaches, at least for moderately large $n$. From a closer look at equation~\eqref{eq:pi_modpost} we see that the only variables changing with $k$ are $\lambda_k$ and $r_k$.  If we cache $\bepsilon_{:,C}$, the only costly operation in it is the modified Bessel function. We can approximate equation~\eqref{eq:pi_modpost} by
\begin{align} \label{eq:pi_fmodpost}
	Z_{k,C} \ \propto & \ \frac{K_{1-d/2}(\sqrt{\bepsilon_{k-1,C}})}{(\bepsilon_{k-1,C}\inv)^{(1-d/2)/2}}\exp\left(\frac{\lambda_k}{2} r_k\right) \,,
\end{align}
where we have dropped $\lambda_k$ from the Bessel function and the divisor in equation~\eqref{eq:pi_modpost}. This is acceptable because (i) $K_{\nu}(z)$ is strictly decreasing for fixed $\nu$ and (ii) the $\lambda_k$ term appearing in the divisor is a constant in $\log Z_{k,C}$. Note that a similar reasoning can be applied to equation~\eqref{eq:pi_modpost_ne}, which is cheaper and numerically more stable. Since equation~\eqref{eq:pi_fmodpost} depends on $k$ only through $\lambda_k r_k$, we have decreased the cost from ${\cal O}(\kappa_2n^3 +\kappa_3n)$ to ${\cal O}(\kappa_2n^2 + \kappa_3n)$, that is, we need to compute equation~\eqref{eq:pi_fmodpost} for every possible pair only once before selecting the merging pair at stage $k$, then we add $\lambda_k r_k/2$ (in log-domain) before sampling its merging time. From now on we use {\sc MPost1} to refer to the algorithm using equation \eqref{eq:pi_modpost} and {\sc MPost2} to the fast approximation in equation~\eqref{eq:pi_fmodpost}.

Recently, \citet{gorur08} proposed an efficient SMC sampler ({\sc SMC1}) for hierarchical clustering with coalescents that although was introduced for discrete data can be easily adapted to continuous data. Their approach is based on a regenerative race process in which every possible pair candidate proposes a merging time only once leading to ${\cal O}(\kappa_1n^2)$ computational time. In principle, {\sc SMC1} has quadratic cost as does {\sc MPost2}, however since $\kappa_2<<\kappa_1$ our approximation is considerably faster. In addition, we have observed empirically that at least for $n$ in the lower hundreds, {\sc MPost1} is also faster than {\sc SMC1}.

The key difference between our approach and that of \citet{gorur08} is that the latter proposes merging times for every possible pair and selects the pair to merge as the minimum available at a given stage whereas our approach selects the pair to merge and samples the merging time in a separate step. Additionally, they do not sample merging times using $\Delta_{k,c}=t_k-t_c$ but directly $t_k|t_c$, where $t_c$ is the time at which the pair $c$ was created, thus $0\leq t_c < t_k$. For example all pairs of singletons get created at $t_c=0$ regardless of the value of $k$. As $t_c$ occurs generally earlier than $t_{k-1}$, {\sc SMC1} draws $\Delta_{k,c}$ in larger jumps compared to {\sc MPost1/2}. Since the conditioning of $t_k$ is involved in the truncation level of the GIG distribution, time samples from {\sc SMC1} are less constrained. This suggests that our approach will have in general better mixing properties; we compare the speed and accuracy of our approach with previous approaches on both simulated and real data examples in Appendix \ref{sc:results}.

\section{Application: Multiple correlated outcomes}\label{sec:emrResults}
The broad adoption of electronic medical records has created the possibility of building predictive models for pateint populations at almost any health system.  One of the challenges faced when building these predictive models is identifying the appropriate outcome.  This is because patients with chronic diseases are typically at risk for many different bad outcomes - often with varying levels of relatedness.  In this section we will demostrate the incorporation of the Kingman coalescent in a larger model that allows one to borrow strength across multiple outcomes with unknown levels of relatedness.

Suppose we have $Y$, an $N\times P$ dimensional matrix with elements $y_{ij}$ indicating the presence or absence of outcome $j$ for patient $i$.  Also, let $X$ be a $N\times K$ dimensional matrix of independent variables.  We assume a probit regression model for the outcomes.  For $x_i$, the $i^{th}$ row of $X$ we have:
\begin{eqnarray*}
P(y_{ij}) &=& \Psi(\mu_j + x_i \beta_j)
\end{eqnarray*}
where the outcome specific intercept $\mu_j$ allows joint modeling of outcomes with different rates, $\beta_j$ is the $K$ dimensional vector of regression coefficients for outcome $j$ and $\Psi$ is the cumulative distribution function for the standard normal distribution.  Because all outcomes are at increased risk from the same chronic disease, we suspect that probabilites will be correlated.  In order to capture this, we use the coalescent to impose a prior distribution that encourages correlation in the regression coefficients without {\em a priori} knowledge of the strength of relatedness or the values of the regression coefficients.  Suppose $\beta_k$ is the parent node for $\beta_j$ and recall that $t_k$ is merging time for node $k$. 
\begin{eqnarray*}
\beta_j|\beta_k, \t, \bpi &=& N(\beta_k,(t_k-t_j) \Phi)\\
\t, \bpi &\sim & \mbox{Coalescent}(K)
\end{eqnarray*}
We use an inverse Wishart prior on the covariance matrix $\Phi$.

We will test the model in a factor regression context examining models that predict outcomes for $\approx$ 19K diabetic patients using data from an electronic medical record.  Diabetic patients are susceptible to many different types of comorbidities - we develop predictive models for the 21 that are listed in Table \ref{tbl:results}.  The presence/absence of these comorbidities was ascertained through ICD9 codes; the codes used to identify each are available in supplementary material. 

We conducted our experiment on data from a large regional health system - Duke University IRB protocol number PRO00060340.  The patient pool consisted of all patients with a home address in a particular county and all records collected between 2007 and 2011 (inclusive).  

{\em Construction of independent variables and outcomes.}  For each patient, the data consists of a collection of date-time stamped observations with labels.  The labels include medical codes (ICD9, CPT), vitals, medications and laboratory values.  For some types of observations (labs and vitals) there are additional continuous values, but these were discarded for this analysis.  A high-dimensional sparse vector (3865 independent variables) was constructed for each patient by counting observations for 6 months prior to a specified {\em threshold date} - 1/1/2009 for training and 1/1/2010 for test.  We used non-negative matrix factorization \citep{nnmf} on the square root of this matrix to reduce the dimension from 3865 to 25; the 25 factor scores vectors were then used in the regression model above.  

We recognize the opportunity to incorporate the dimension reduction into a larger model rather than using this two-step process of dimension reduction followed by regression.  However, that approach would have complicated the comparison to more standard regression techniques.  We also note that the matrix factorization generates collections of observation types with very nice medical interpretations; for example, one factor is heavily weighted with medical codes, meds and labs associated with lung cancer. However, examination of those discovered relationships is outside the scope of this article.

\begin{table}[h]
\scriptsize
\begin {longtable}{r|ccccc|ccccc}
\rowcolor [rgb]{1,1,1}&\multicolumn{5}{c}{\underline{Training data set, 10-fold cross-validation}}&\multicolumn{5}{c}{\underline{Validation data set}}\\
\rowcolor [rgb]{1,1,1}Outcome&\# Events&CR&RF&MLE&Lasso&\# Events&CR&RF&MLE&Lasso\\\hline
\rowcolor [gray]{.9}Death&450&\textbf{0.814}&0.814&0.81&0.791&376&\textbf{0.789}&0.784&0.786&0.761\\ 
\rowcolor [gray]{.8}Acute MI&190&\textbf{0.704}&0.658&0.7&0.657&197&\textbf{0.699}&0.648&0.699&0.694\\
\rowcolor [gray]{.9}Amputation&42&\textbf{0.75}&0.684&0.709&0.443&38&\textbf{0.758}&0.605&0.705&0.504\\ 
\rowcolor [gray]{.8}Aneurysm&25&0.676&\textbf{0.737}&0.64&0.355&16&\textbf{0.735}&0.582&0.706&0.508\\ 
\rowcolor [gray]{.9}Angioplasty&72&\textbf{0.616}&0.59&0.575&0.39&71&\textbf{0.605}&0.544&0.578&0.543\\ 
\rowcolor [gray]{.8}Arterial Cath&82&\textbf{0.759}&0.724&0.737&0.476&101&\textbf{0.673}&0.673&0.653&0.543\\ 
\rowcolor [gray]{.9}Atrial Fibrillation&619&0.817&\textbf{0.828}&0.818&0.809&593&0.799&\textbf{0.816}&0.798&0.798\\ 
\rowcolor [gray]{.8}Bipolar&199&\textbf{0.702}&0.649&0.694&0.603&216&\textbf{0.717}&0.691&0.711&0.548\\ 
\rowcolor [gray]{.9}Cardiac Cath&252&\textbf{0.648}&0.605&0.644&0.477&258&\textbf{0.668}&0.619&0.653&0.545\\ 
\rowcolor [gray]{.8}Skin Ulcer&286&0.727&0.692&\textbf{0.733}&0.655&338&0.705&0.689&\textbf{0.711}&0.673\\ 
\rowcolor [gray]{.9}COPD&296&0.726&\textbf{0.729}&0.728&0.704&294&\textbf{0.721}&0.7&0.718&0.713\\ 
\rowcolor [gray]{.8}Coronary Disease&1439&0.752&\textbf{0.779}&0.752&0.743&1406&0.747&\textbf{0.761}&0.747&0.732\\ 
\rowcolor [gray]{.9}Depression&194&0.715&0.693&\textbf{0.718}&0.707&176&0.733&0.71&0.737&\textbf{0.748}\\ 
\rowcolor [gray]{.8}Heart Failure&470&\textbf{0.806}&0.801&0.806&0.799&556&\textbf{0.785}&0.756&0.785&0.774\\ 
\rowcolor [gray]{.9}Kidney Disease&456&0.781&\textbf{0.807}&0.776&0.768&415&0.765&\textbf{0.787}&0.767&0.748\\ 
\rowcolor [gray]{.8}Neurological&636&\textbf{0.751}&0.745&0.75&0.745&668&0.734&0.717&\textbf{0.736}&0.73\\ 
\rowcolor [gray]{.9}Obesity&1647&0.676&0.675&\textbf{0.677}&0.664&1926&\textbf{0.639}&0.639&0.639&0.625\\ 
\rowcolor [gray]{.8}Opthalmic&491&0.739&\textbf{0.744}&0.738&0.736&480&0.737&\textbf{0.741}&0.739&0.724\\ 
\rowcolor [gray]{.9}Arthritis&57&\textbf{0.616}&0.486&0.601&0.362&62&0.602&0.598&\textbf{0.604}&0.431\\ 
\rowcolor [gray]{.8}Stroke&344&\textbf{0.741}&0.698&0.737&0.727&334&\textbf{0.692}&0.67&0.687&0.68\\ 
\rowcolor [gray]{.9}Unstable Angina&173&\textbf{0.66}&0.616&0.657&0.501&125&0.72&0.689&0.712&\textbf{0.729}
\end {longtable}
\caption{Comparison of predictive performance of different algorithms. CR=coalescent regression, RF=random forest, MLE=maximum likelihood. We also compared elastic net and ridge; because those approaches showed worse performance in all outcomes we omitted them from this table for clarity.  The full table is availabe in supplementary material.}\label{tbl:results}
\end{table}

Outcomes were constructed by identifying the presence/absence of the relevant comorbidity in the year following the threshold date.  Training and test data sets were constructed by thresholding the dataset at Jan 1, 2009 and Jan 1, 2010 respectively.  We compare our results to ridge regression, elastic net and maximum likelihood regression.

{\em Predictive accuracy.} Predictive accuracy was assessed by area under the ROC curve using 10-fold cross-validation in the training set and directly on the test set after training on the full training set.  Table \ref{tbl:results} shows accuracies for each of the outcomes and a number of the modeling methodologies.  The model described here, coalescent regression, consistently outperformed other approaches.  Elastic net and Ridge were also tested, but performed worse than the others listed.  The full table including elastic net and ridge is available in supplemental material.

{\em Future costs for diabetic patients.}  Identifying high risk patients with the goal of preventing morbidity and mortality is of paramount interest.  However, improving the health of the high risk patients can have the additional side effect of lowering their future healthcare costs.  In order to assess the future costs of patients with diabetes, we linked the billing codes from the medical records data set with the publicaly available physician fee schedule from the Centers for Medicare and Medicaid \citep{cmsFee}.  This allows us to assign a cost to each procedure performed for each patient.  We then computed the cumulative cost at each day following the threshold date for each patient.

\begin{figure}[!ht]
\captionsetup[subfigure]{}
\centering
\subfloat[]{\includegraphics[width=.45\textwidth]{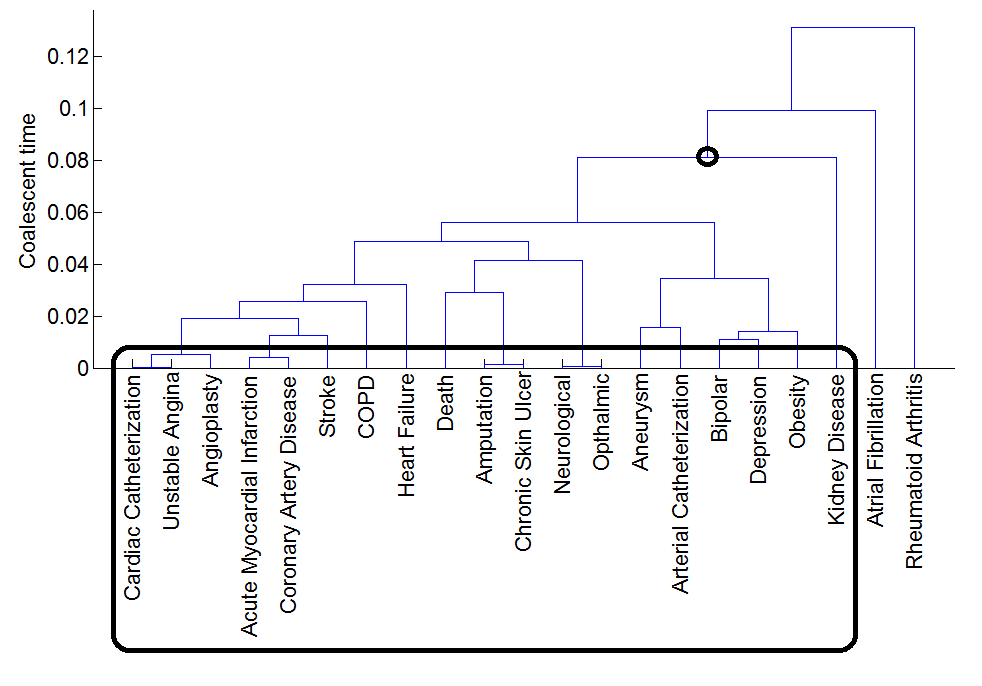}}\quad
\subfloat[]{\includegraphics[width=.45\textwidth]{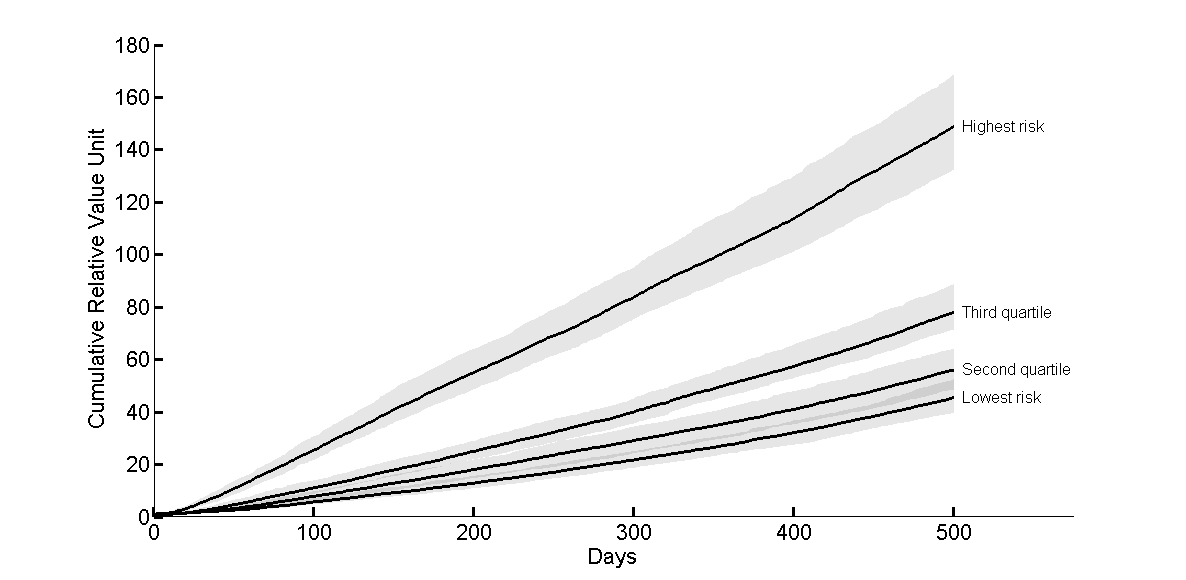}}
\caption{Future costs for patients with diabetes.  Panel (a) shows the tree structure with highest posterior probability; the circled node corresponds to the model that best stratifies patients on future costs in the {\em training} data set.  Panel (b) shows future costs of patients in the {\em test} data set stratified by quartile from the risk model identified in panel (a).  90\% confidence bands are computed by bootstrapping with 500 patients.} 
\label{fig:futureCost} 
\end{figure}

One advantage of the tree model for predicting outcomes is the availability of regression models at each of the parent nodes.  Models in these nodes can be loosely interpreted as predictors of the composite outcome composed of outcomes in each of their associated leaf nodes.  Each of the leaf and internal node models gives rise to a separate risk stratification.  We tested - in the training data set - the ability of all models to stratify patients based on future costs.  Figure \ref{fig:futureCost}(a) shows the tree with the highest posterior probability among those sampled; the circled node corresponds to the model that best stratifies patients based on future cost in the {\em training} data set.  Costs for each of four quartiles based on risk stratification using this model are shown in the {\em test} data set in Figure \ref{fig:futureCost}(b).  Because the outcomes we are predicting with our risk model are typically high cost outcomes, we find the resulting models stratify future costs very well even though future cost was not explicitly included in the model.

%
%
%
\section*{Supplementary materials}
{\bf Source code:} Matlab/C code required to perform the methods described in the manuscript, including sample scripts and the artificial data generator used throughout the results section (mvbtree\_v0.1.zip, compressed file).
%
%
\appendix
\section{Tree posteriors details} \label{eq:delta_pi_details}
Recall that $v_k=2\Delta_k+r_k$, $\lambda_k=(n-k+1)(n-k)/2$, and $\bepsilon_{k-1,C}=(\m_{c_1}-\m_{c_2})\bPhi\inv(\m_{c_1}-\m_{c_2})\ts$. We can combine equations \eqref{eq:ZkNorm} and \eqref{eq:delta_pi_post} to get:
\begin{align}
p(\Delta_k,\pi_{k-1,C}|\pi_{1:k-1}, \t_{1:k-1})& =  \ \DN\left(\m_{c_1}-\m_{c_2}\middle|\0,v_k\bPhi\right)\exp\left(-v_k\middle|\lambda_k/2\right)\exp(r_k\lambda_k/2) \,, \nonumber \\
	& = \frac{\lambda_k}{2}(2\pi)^{-d/2}|\bPhi|^{-1/2}\underbrace{|v_k|^{-d/2}\exp\left(-\frac{\bepsilon_{k-1,C}}{2}v_k\inv-\frac{\lambda_k}{2}v_k\right)}\exp\left(\frac{\lambda_k}{2}r_k\right) \,, \label{eq:gig_core}
\end{align}
We recognize the segment in braces of equation~\eqref{eq:gig_core} as the core of a generalized inverse Gaussian distribution \citep{jorgensen82a} defined as
%
\begin{align*}
	{\rm GIG}(x|\lambda_k,\chi,\psi) \ = \ 	\frac{(\psi\chi\inv)^{(1-d/2)/2}}{2K_{1-d/2}(\sqrt{\psi\chi})}x^{\lambda_k-1}\exp\left(-\frac{\chi}{2}x\inv-\frac{\psi}{2}x\right) \,,
\end{align*}
where $K_\nu(z)$ is the modified Bessel function of second kind with parameter $\nu$.

We can thus use equation~\eqref{eq:gig_core} to obtain
\begin{align*}
	\Delta_k|\pi_{1:k},\t_{1:k-1} \ \sim & \ \tGIG_{r_k}(v_k|\bepsilon_{k,C},\lambda_k) \,, \\
	\pi_k|\t_{1:k-1},\pi_{1:k-1}  \ \propto & \ \underbrace{\frac{\lambda_k^2}{4}(2\pi)^{-d/2}\frac{2K_{1-d/2}(\sqrt{\lambda_k\bepsilon_{k-1,C}})}{(\lambda_k\bepsilon_{k-1,C}\inv)^{(1-d/2)/2}}|\bPhi|^{-1/2}\exp(\frac{\lambda_k}{2}r_k)}_{Z_{k,C}} \,, \\
	\pi_k|\t_{1:k-1},\pi_{k-1}  \ \leq & \ h\underbrace{\frac{K_{1-d/2}(\sqrt{\lambda_k\bepsilon_{k-1,C}})}{(\lambda_k\bepsilon_{k-1,C}\inv)^{(1-d/2)/2}}|\bPhi|^{-1/2}\exp(\frac{\lambda_k}{2}r_k)}_{Z_{k,C}} \,,
\end{align*}
where the distribution for $\Delta_k$ is truncated below $r_k$ because of the constraint $v_k=2\Delta_k+r_k$, with $r_k>0$, $h$ is a constant term and the inequality is the result of not being able to obtain a closed form for the normalization constant of the truncated generalized inverse Gaussian distribution.
%
%
\FloatBarrier

\begin{sidewaystable}
\section{Full results table from Section \ref{sec:emrResults}}
\small
\begin {longtable}{rccccccc|ccccccc}
\rowcolor [rgb]{1,1,1}&\multicolumn{7}{c}{\underline{Training data set, 10-fold cross-validation}}&\multicolumn{7}{c}{\underline{Validation data set}}\\
\rowcolor [rgb]{1,1,1}Outcome&\# Events&CR&RF&MLE&Lasso&Enet&Ridge&\# Events&CR&RF&MLE&Lasso&Enet&Ridge\\\hline
\rowcolor [gray]{.9}Death&450&\textbf{0.814}&0.814&0.81&0.791&0.79&0.794&376&\textbf{0.789}&0.784&0.786&0.761&0.761&0.762\\ 
\rowcolor [gray]{.8}Acute MI&190&\textbf{0.704}&0.658&0.7&0.657&0.673&0.676&197&\textbf{0.699}&0.648&0.699&0.694&0.539&0.697\\
\rowcolor [gray]{.9}Amputation&42&\textbf{0.75}&0.684&0.709&0.443&0.356&0.445&38&\textbf{0.758}&0.605&0.705&0.504&0.504&0.504\\ 
\rowcolor [gray]{.8}Aneurysm&25&0.676&\textbf{0.737}&0.64&0.355&0.335&0.382&16&\textbf{0.735}&0.582&0.706&0.508&0.508&0.508\\ 
\rowcolor [gray]{.9}Angioplasty&72&\textbf{0.616}&0.59&0.575&0.39&0.415&0.406&71&\textbf{0.605}&0.544&0.578&0.543&0.543&0.543\\ 
\rowcolor [gray]{.8}Arterial Cath&82&\textbf{0.759}&0.724&0.737&0.476&0.517&0.568&101&\textbf{0.673}&0.673&0.653&0.543&0.543&0.543\\ 
\rowcolor [gray]{.9}Atrial Fibrillation&619&0.817&\textbf{0.828}&0.818&0.809&0.814&0.814&593&0.799&\textbf{0.816}&0.798&0.798&0.793&0.793\\ 
\rowcolor [gray]{.8}Bipolar&199&\textbf{0.702}&0.649&0.694&0.603&0.579&0.617&216&\textbf{0.717}&0.691&0.711&0.548&0.668&0.704\\ 
\rowcolor [gray]{.9}Cardiac Cath&252&\textbf{0.648}&0.605&0.644&0.477&0.504&0.503&258&\textbf{0.668}&0.619&0.653&0.545&0.545&0.545\\ 
\rowcolor [gray]{.8}Skin Ulcer&286&0.727&0.692&\textbf{0.733}&0.655&0.629&0.66&338&0.705&0.689&\textbf{0.711}&0.673&0.659&0.664\\ 
\rowcolor [gray]{.9}COPD&296&0.726&\textbf{0.729}&0.728&0.704&0.689&0.693&294&\textbf{0.721}&0.7&0.718&0.713&0.701&0.712\\ 
\rowcolor [gray]{.8}Coronary Disease&1439&0.752&\textbf{0.779}&0.752&0.743&0.743&0.743&1406&0.747&\textbf{0.761}&0.747&0.732&0.732&0.736\\ 
\rowcolor [gray]{.9}Depression&194&0.715&0.693&\textbf{0.718}&0.707&0.67&0.709&176&0.733&0.71&0.737&\textbf{0.748}&0.748&0.746\\ 
\rowcolor [gray]{.8}Heart Failure&470&\textbf{0.806}&0.801&0.806&0.799&0.798&0.8&556&\textbf{0.785}&0.756&0.785&0.774&0.773&0.773\\ 
\rowcolor [gray]{.9}Kidney Disease&456&0.781&\textbf{0.807}&0.776&0.768&0.768&0.768&415&0.765&\textbf{0.787}&0.767&0.748&0.743&0.746\\ 
\rowcolor [gray]{.8}Neurological&636&\textbf{0.751}&0.745&0.75&0.745&0.746&0.747&668&0.734&0.717&\textbf{0.736}&0.73&0.73&0.731\\ 
\rowcolor [gray]{.9}Obesity&1647&0.676&0.675&\textbf{0.677}&0.664&0.666&0.669&1926&\textbf{0.639}&0.639&0.639&0.625&0.629&0.63\\ 
\rowcolor [gray]{.8}Opthalmic&491&0.739&\textbf{0.744}&0.738&0.736&0.734&0.733&480&0.737&\textbf{0.741}&0.739&0.724&0.724&0.724\\ 
\rowcolor [gray]{.9}Arthritis&57&\textbf{0.616}&0.486&0.601&0.362&0.405&0.394&62&0.602&0.598&\textbf{0.604}&0.431&0.431&0.431\\ 
\rowcolor [gray]{.8}Stroke&344&\textbf{0.741}&0.698&0.737&0.727&0.725&0.725&334&\textbf{0.692}&0.67&0.687&0.68&0.68&0.683\\ 
\rowcolor [gray]{.9}Unstable Angina&173&\textbf{0.66}&0.616&0.657&0.501&0.484&0.561&125&0.72&0.689&0.712&\textbf{0.729}&0.584&0.726
\end {longtable}
\end{sidewaystable}
\FloatBarrier

\section{Numerical results} \label{sc:results}
In this section we consider a number of experiments on both artificial and real data to highlight the benefits of the proposed approaches as well as to compare them with previously proposed ones. In total three artificial data based simulations and two real data set applications are presented. All experiments are obtained using a desktop machine with 2.8GHz processor with 8GB RAM and run times are measured as single core CPU times. All methods but standard hierarchical clustering were coded by the authors using Matlab and C in order to make run times a fair proxy measurement to computational cost.
\subsection{Artificial data - structure}
First we compare different sampling algorithms on artificially generated data using $n$-coalescents and Gaussian processes with known squared exponential covariance functions as priors. We generated 50 replicates of two different settings, $D_1$ and $D_2$ of sizes $\{n,d\}=\{32,32\}$ and $\{64,64\}$, respectively. We compare four different algorithms, {\sc Post-Post} \citep{teh08}, {\sc SCM1} \citep{gorur08}, {\sc MPost1} and {\sc MPost2}. In each case we collect $M=100$ particles and set the covariance function parameters $\{\ell,\sigma^2\}$ to their true values. As performance measures we track runtime ({\sc rt}) as proxy to the computational cost, mean squared error ({\sc mse}), mean absolute error ({\sc mae}) and maximum absolute bias ({\sc mab}) of $\log(\t)$ and $\bpi$. 

For $\bpi$ we compute the distance matrix encoded by $\{\t,\bpi\}$. In addition, we compute the difference between true and estimated last coalescing times ({\sc td}) as a way to quantify estimation bias, thus positive and negative differences imply under and over estimation, respectively. 

Table~\ref{tb:artificial_structure} shows performance measures averaged over 50 replicates for each data set. In terms of error, we see that all four algorithms perform about the same as one might expect, however with {\sc MPost1} and {\sc SMC1} being slightly better and slightly worse, respectively. The computational cost is significantly higher for the {\sc Post-Post} approach whereas {\sc MPost2} is the fastest. We see {\sc MPost1} and {\sc MPost2} consistently outperforming the other two algorithms as an indication of better mixing properties. In more general terms, {\sc MPost2} provides the best error/computational cost trade-off as the difference in accuracy between {\sc MPost1} and {\sc MPost2} is rather minimal.  In terms of coalescing time estimation bias, we see that {\sc Post-Post} performs slightly better and {\sc SMC1} has the largest bias from all four. Note that {\sc Post-Post} is not affected by the approximation introduced in equation~\eqref{eq:pi_modpost} hence it should be in principle unbiased.
\begin{table}[!t]
	\centering\scriptsize{
	\begin{tabular}{cccccc}
	\hline
	Set & Measure & {\sc Post-post} & {\sc MPost1} & {\sc SMC1} & {\sc MPost2} \\
	\hline
	\multicolumn{6}{l}{Merge time ($\t$)} \\
	\multirow{3}{*}{$D_1$} & $10^{1}\times${\sc mse} & $0.52\pm0.22$ & $\bf 0.44\pm0.18$ & $0.66\pm0.28$ & $0.45\pm0.18$ \\
	& $10^{1}\times${\sc mae} & $1.88\pm0.45$ & $\bf 1.68\pm0.39$ & $2.01\pm0.46$ & $1.72\pm0.40$ \\
	& $10^{1}\times${\sc mab} & $5.13\pm1.23$ & $4.93\pm1.12$ & $6.09\pm1.30$ & $\bf 4.91\pm1.08$ \\
	\multirow{3}{*}{$D_2$} & $10^{2}\times${\sc mse} & $4.06\pm1.08$ & $\bf 3.04\pm0.90$ & $5.37\pm2.23$ & $3.30\pm0.94$ \\
	& $10^{1}\times${\sc mae} & $1.73\pm0.26$ & $\bf 1.42\pm0.26$ & $1.83\pm0.39$ & $1.49\pm0.27$ \\
	& $10^{1}\times${\sc mab} & $4.47\pm0.73$ & $4.40\pm0.78$ & $5.97\pm1.39$ & $\bf 4.39\pm0.73$ \\
	\multicolumn{6}{l}{Last coalescing time ($t_{n-1}$)} \\
	\multirow{1}{*}{$D_1$} & $10^{0}\times${\sc td} & $\bf -0.04\pm0.52$ & $-0.05\pm0.52$ & $0.11\pm0.55$ & $-0.07\pm0.51$ \\
	\multirow{1}{*}{$D_2$} & $10^{0}\times${\sc td} & $\bf -0.12\pm0.51$ & $-0.13\pm0.50$ & $0.13\pm0.49$ & $-0.13\pm0.63$ \\
	\multicolumn{6}{l}{Distance matrix ($\bpi$)} \\
	\multirow{3}{*}{$D_1$} & $10^{1}\times${\sc mse} & $0.79\pm0.50$ & $0.70\pm0.49$ & $1.32\pm0.63$ & $\bf 0.70\pm0.42$ \\
	& $10^{1}\times${\sc mae} & $2.24\pm0.78$ & $\bf 2.13\pm0.76$ & $3.03\pm0.90$ & $2.14\pm0.72$ \\
	& $10^{1}\times${\sc mab} & $6.77\pm1.20$ & $6.50\pm1.12$ & $8.77\pm1.96$ & $\bf 6.48\pm1.13$ \\
	\multirow{3}{*}{$D_2$} & $10^{2}\times${\sc mse} & $5.79\pm3.15$ & $\bf 4.89\pm2.92$ & $11.36\pm5.68$ & $5.27\pm2.94$ \\
	& $10^{1}\times${\sc mae} & $1.95\pm0.68$ & $\bf 1.78\pm0.65$ & $2.85\pm0.83$ & $1.85\pm0.65$ \\
	& $10^{1}\times${\sc mab} & $6.06\pm0.79$ & $\bf 5.75\pm0.73$ & $8.54\pm2.01$ & $5.81\pm0.69$ \\
	\multicolumn{6}{l}{Computational cost} \\
	\multirow{1}{*}{$D_1$} & $10^{0}\times${\sc rt} & $18.65\pm0.24$ & $2.29\pm0.04$ & $3.76\pm0.07$ & $\bf 1.98\pm0.03$ \\
	\multirow{1}{*}{$D_2$} & $10^{-1}\times${\sc rt} & $14.50\pm0.06$ & $1.08\pm0.00$ & $1.39\pm0.01$ & $\bf 0.61\pm0.00$ \\
	\hline
	\end{tabular}}
	\caption[Performance measures for structure estimation]{Performance measures for structure estimation. {\sc mse}, {\sc mae}, {\sc mab}, {\sc rt} and {\sc td} are mean squared error, mean absolute error, maximum absolute bias, runtime in seconds and last coalescing time difference, respectively. Figures are means and standard deviations across 50 replicates. Best results are in boldface letters.}
	\label{tb:artificial_structure}
\end{table}

\subsection{Artificial data - covariance}
Next we want to test the different sampling algorithms when the parameters of the Gaussian process covariance function, $\{\ell,\sigma^2\}$, need to be learned as well. We use settings similar to those in the previous experiment with the difference that now $M=50$ particles are collected and $N_{\rm iter}=50$ iterations are performed to learn the covariance matrix parameters. We dropped the first 10 iterations as burn-in period. In addition to the previously mentioned performance measures we also compute {\sc mse}, {\sc mae} and {\sc mab} for the inverse length scale $\ell$ from equation~\eqref{eq:sqexp_noise}. In order to simplify the experiment we set a priori $\sigma^2=1\times10^{-9}$ to match a \emph{noiseless} scenario, however similar results are obtained when learning both parameters at the same time (results not shown). Table~\ref{tb:artificial_covariance} shows an overall similar trend when compared to Table~\ref{tb:artificial_structure}. In terms of covariance function parameter estimation, we see all algorithms perform about the same which is not surprising considering they use the same sampling strategy. Time difference measures, {\sc td}, indicate that all approaches underestimate the last coalescing time, although slightly less for our two proposals. More specifically, median differences between {\sc SMC1} and {\sc MPost1/2} are significant at the 0.05 level only for the distance matrix measure, which indicates that our algorithms are better at capturing the true hierarchical structure of the data but all methods do about the same in terms of coalescing times and inverse length-scale estimation. {\sc MPost1/2} is approximately twice as fast as {\sc SMC1}. Other than computational speed, differences between {\sc MPost1} and {\sc MPost2} are not significant.
\begin{table}[!t]
	\centering\scriptsize{
	\begin{tabular}{ccccc}
	\hline
	Set & Measure & {\sc MPost1} & {\sc SMC1} & {\sc MPost2} \\
	\hline
	\multicolumn{5}{l}{Merge time ($\t$)} \\
	\multirow{3}{*}{$D_1$} & $10^{1}\times${\sc mse} & $\bf 1.20\pm0.51$ & $1.26\pm0.44$ & $1.24\pm0.53$ \\
	& $10^{1}\times${\sc mae} & $3.02\pm0.80$ & $\bf 2.98\pm0.62$ & $3.06\pm0.81$ \\
	& $10^{1}\times${\sc mab} & $\bf 6.43\pm1.26$ & $7.01\pm1.50$ & $6.52\pm1.38$ \\
	\multirow{3}{*}{$D_2$} & $10^{2}\times${\sc mse} & $\bf 5.38\pm1.66$ & $6.32\pm1.94$ & $5.61\pm1.76$ \\
	& $10^{1}\times${\sc mae} & $2.02\pm0.36$ & $\bf 2.01\pm0.36$ & $2.07\pm0.36$ \\
	& $10^{1}\times${\sc mab} & $4.75\pm0.72$ & $6.16\pm1.35$ & $\bf 4.73\pm0.62$ \\
	\multicolumn{5}{l}{Last coalescing time ($t_{n-1}$)} \\
	\multirow{1}{*}{$D_1$} & $10^{0}\times${\sc td} & $\bf 0.21\pm0.52$ & $0.43\pm0.51$ & $0.25\pm0.55$ \\
	\multirow{1}{*}{$D_2$} & $10^{0}\times${\sc td} & $\bf 0.04\pm0.44$ & $0.29\pm0.54$ & $0.08\pm0.35$ \\
	\multicolumn{5}{l}{Distance matrix ($\bpi$)} \\
	\multirow{3}{*}{$D_1$} & $10^{1}\times${\sc mse} & $\bf 1.31\pm0.87$ & $2.85\pm1.76$ & $1.33\pm0.86$ \\
	& $10^{1}\times${\sc mae} & $\bf 2.94\pm1.19$ & $4.53\pm1.69$ & $2.96\pm1.19$ \\
	& $10^{1}\times${\sc mab} & $\bf 8.77\pm2.18$ & $9.60\pm1.73$ & $8.84\pm2.13$ \\
	\multirow{3}{*}{$D_2$} & $10^{1}\times${\sc mse} & $\bf 0.64\pm0.35$ & $1.54\pm0.67$ & $0.66\pm0.34$ \\
	& $10^{1}\times${\sc mae} & $\bf 2.08\pm0.63$ & $3.29\pm0.94$ & $2.08\pm0.65$ \\
	& $10^{1}\times${\sc mab} & $6.77\pm1.13$ & $8.41\pm1.42$ & $\bf 6.76\pm1.15$ \\
	\multicolumn{5}{l}{Inverse length scale ($\ell$)} \\
	\multirow{3}{*}{$D_1$} & $10^{4}\times${\sc mse} & $\bf 2.36\pm3.20$ & $2.93\pm4.51$ & $2.38\pm3.23$ \\
	& $10^{2}\times${\sc mae} & $\bf 1.17\pm0.99$ & $1.24\pm1.09$ & $1.18\pm0.99$ \\
	& $10^{2}\times${\sc mab} & $1.40\pm1.14$ & $2.06\pm2.18$ & $\bf 1.38\pm1.15$ \\
	\multirow{3}{*}{$D_2$} & $10^{4}\times${\sc mse} & $2.86\pm3.22$ & $3.55\pm4.48$ & $\bf 2.83\pm3.17$ \\
	& $10^{2}\times${\sc mae} & $1.35\pm1.02$ & $1.44\pm1.14$ & $\bf 1.34\pm1.01$ \\
	& $10^{2}\times${\sc mab} & $1.57\pm1.29$ & $2.39\pm2.24$ & $\bf 1.55\pm1.27$ \\
	\multicolumn{3}{l}{Computational cost} & \\
	\multirow{1}{*}{$D_1$} & $10^{-1}\times${\sc rt} & $5.69\pm0.02$ & $13.65\pm0.06$ & $\bf 4.86\pm0.02$ \\
	\multirow{1}{*}{$D_2$} & $10^{-2}\times${\sc rt} & $2.72\pm0.07$ & $5.12\pm0.05$ & $\bf 1.49\pm0.01$ \\
	\hline
	\end{tabular}}
	\vspace{1mm}
	\caption[Performance measures for covariance estimation]{Performance measures for covariance estimation. {\sc mse}, {\sc mae}, {\sc mab}, {\sc rt} and {\sc dt} are mean squared error, mean absolute error, maximum absolute bias runtime in seconds and last coalescing time difference, respectively. Figures are means and standard deviations across 50 replicates. Best results are in boldface letters.}
	\label{tb:artificial_covariance}
\end{table}

\subsection{Artificial data - greedy algorithm}
As final simulation based on artificial data we want to test wether there is a difference in performance between the mean based greedy approach ({\sc MGreedy}) and the algorithm proposed by \citet{teh08} that utilizes modes ({\sc Greedy}). We generated 50 replicates of two different settings  $D_1$ and $D_2$ of sizes $\{n,d\}=\{32,32\}$ and $\{128,128\}$, respectively. We run $N_{\rm iter}$ iterations of the algorithm and drop the first 10 samples as burn-in period. Other settings and performance measures are the same as in the previous experiment. Table~\ref{tb:artificial_greedy} shows that the algorithm based on means performs consistently better than the original when the data set is small.
\begin{table}[!t]
	\centering\scriptsize
	\begin{tabular}{cccccc}
		\hline
		\multicolumn{3}{c}{$D_1$} & \multicolumn{3}{c}{$D_2$} \\
		Measure & {\sc Greedy} & {\sc MGreedy} & Measure & {\sc Greedy} & {\sc MGreedy} \\
		\hline
		\multicolumn{6}{c}{Merge time ($\t$)} \\
		$10^{1}\times${\sc mse} & $1.17\pm0.46$ & $\bf 0.55\pm0.25$ & $10^{2}\times${\sc mse} & $0.94\pm0.23$ & $\bf 0.59\pm0.15$ \\
		$10^{1}\times${\sc mae} & $2.99\pm0.72$ & $\bf 1.91\pm0.50$ & $10^{1}\times${\sc mae} & $0.80\pm0.11$ & $\bf 0.61\pm0.08$ \\
		$10^{1}\times${\sc mab} & $6.70\pm1.49$ & $\bf 5.46\pm1.45$ & $10^{1}\times${\sc mab} & $2.72\pm0.49$ & $\bf 2.43\pm0.46$ \\
		\multicolumn{6}{c}{Last coalescing time ($t_{n-1}$)} \\
		$10^{0}\times${\sc td} & $0.40\pm0.40$ & $\bf 0.24\pm0.39$ & $10^{0}\times${\sc td} & $0.11\pm0.19$ & $\bf 0.07\pm0.19$ \\
		\multicolumn{6}{c}{Distance matrix ($\bpi$)} \\
		$10^{1}\times${\sc mse} & $1.00\pm0.78$ & $\bf 0.59\pm0.50$ & $10^{1}\times${\sc mse} & $0.11\pm0.11$ & $\bf 0.08\pm0.08$ \\
		$10^{1}\times${\sc mae} & $2.74\pm1.07$ & $\bf 2.04\pm0.88$ & $10^{1}\times${\sc mae} & $0.91\pm0.38$ & $\bf 0.76\pm0.33$ \\
		$10^{1}\times${\sc mab} & $8.69\pm1.56$ & $\bf 7.26\pm1.49$ & $10^{1}\times${\sc mab} & $3.91\pm0.69$ & $\bf 3.59\pm0.66$ \\
		\multicolumn{6}{c}{Inverse length scale ($\ell$)} \\
		$10^{4}\times${\sc mse} & $\bf 0.29\pm2.94$ & $0.30\pm2.92$ & $10^{4}\times${\sc mse} & $\bf 1.37\pm3.54$ & $1.38\pm3.56$ \\
		$10^{2}\times${\sc mae} & $\bf 0.54\pm0.96$ & $\bf 0.54\pm0.96$ & $10^{2}\times${\sc mae} & $\bf 1.17\pm1.03$ & $1.18\pm1.03$ \\
		$10^{2}\times${\sc mab} & $\bf 0.62\pm1.07$ & $0.63\pm1.05$ & $10^{2}\times${\sc mab} & $\bf 1.23\pm1.06$ & $1.24\pm1.15$ \\
		\multicolumn{6}{c}{Computational cost} \\
		$10^{0}\times${\sc rt} & $2.14\pm0.13$ & $2.16\pm0.16$ & $10^{-1}\times${\sc rt} & $\bf 3.04\pm0.14$ & $3.11\pm0.15$ \\
		\hline
	\end{tabular}
	\vspace{1mm}
	\caption[Performance measures for greedy algorithms]{Performance measures for greedy algorithms. {\sc mse}, {\sc mae}, {\sc mab}, {\sc rt} and {\sc dt} are mean squared error, mean absolute error, maximum absolute bias runtime in seconds and last coalescing time difference, respectively. Figures are means and standard deviations across 50 replicates. Best results are in boldface letters.}
	\label{tb:artificial_greedy}
\end{table}

\begin{figure}[!t]
	\centering
	\includegraphics[width=2.5in]{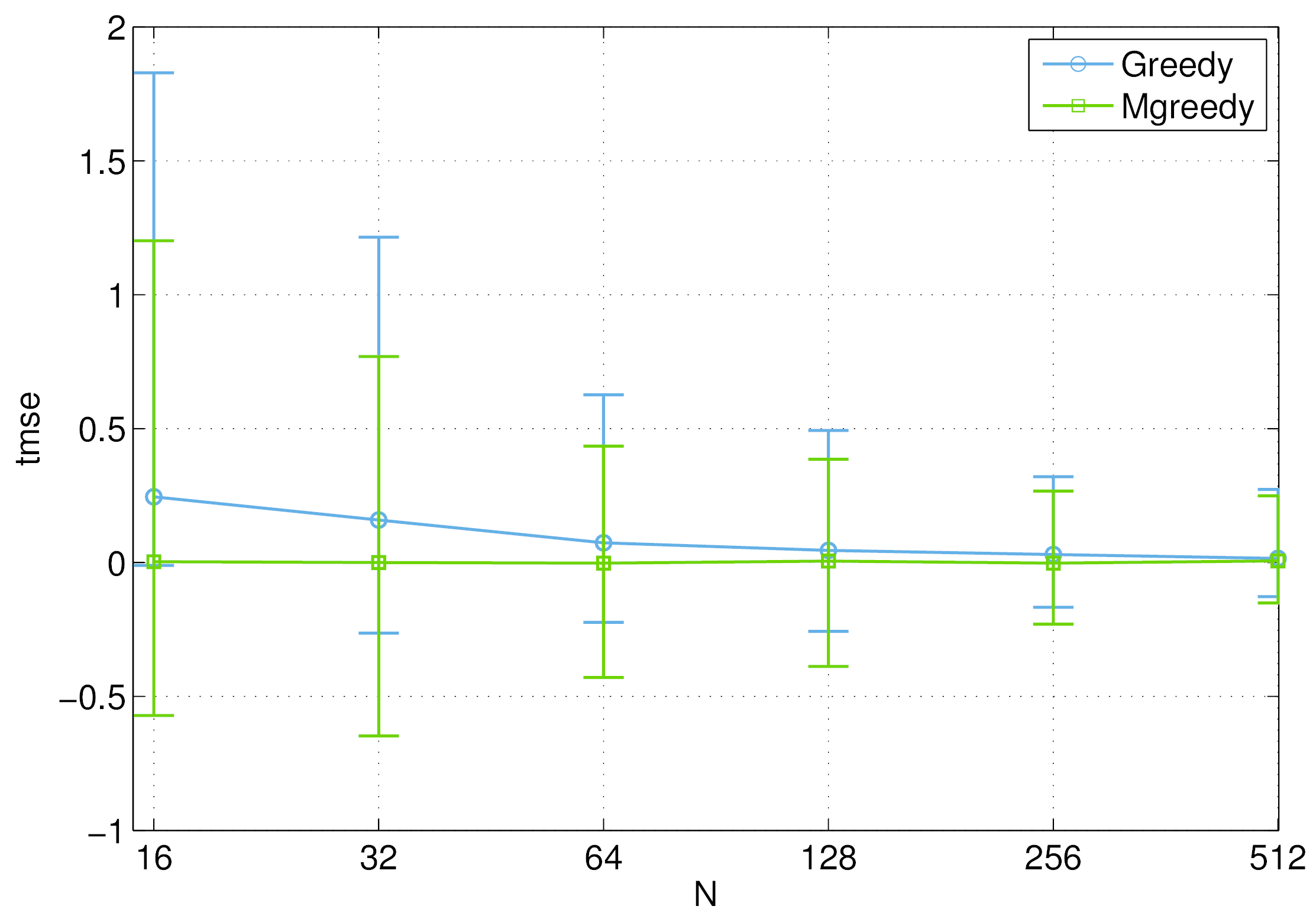}
	\includegraphics[width=2.5in]{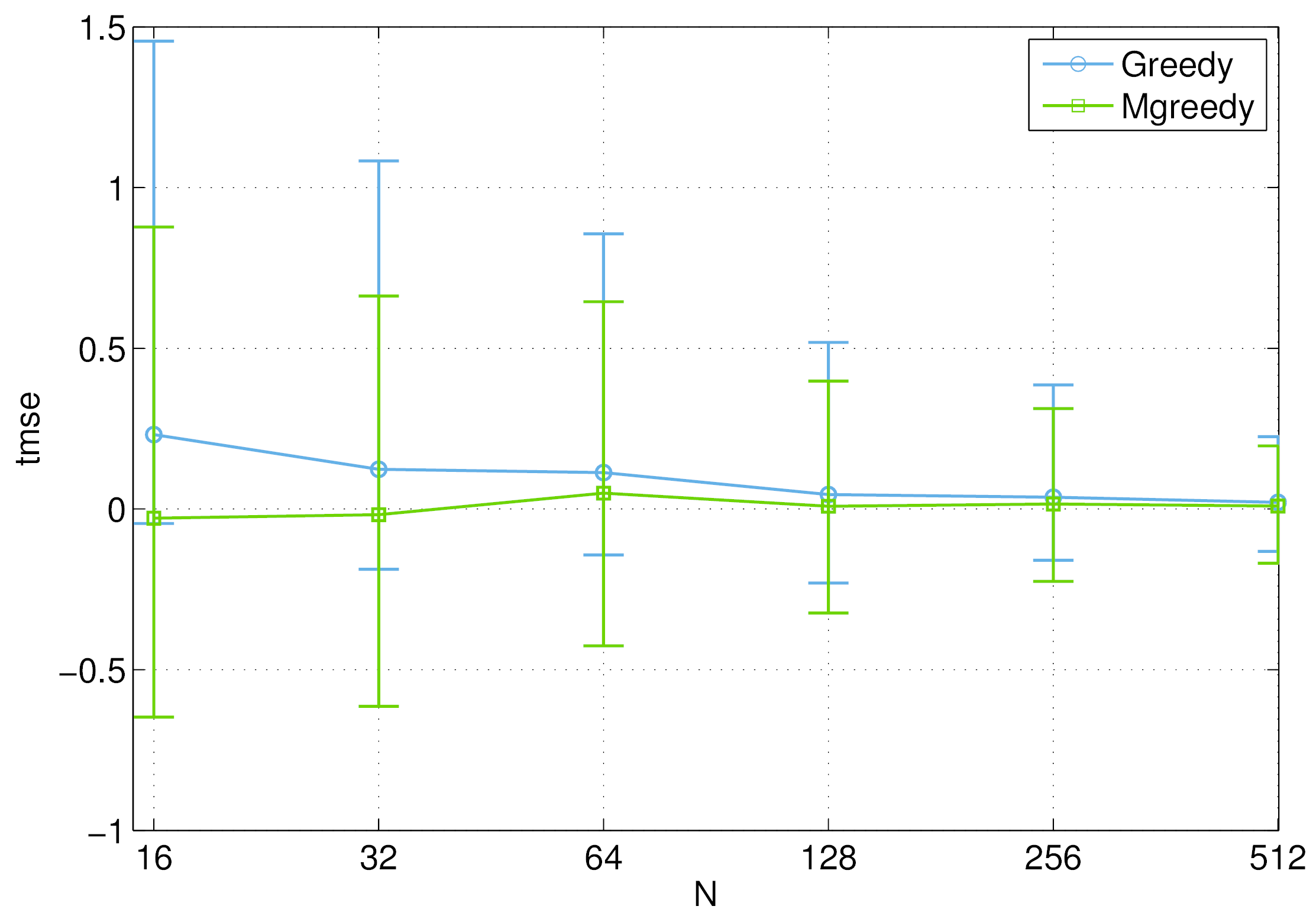}
	\caption[Last coalescing time comparison for greedy algorithms]{Last coalescing time comparison for greedy algorithms. Median time differences {\sc td} as a function of data set dimensionality $d$, for $D_1$ (left) and $D_2$ (right). Error bars represent 90\% empirical quantiles.}
	\label{fg:toy4}
\end{figure}

Although not shown, we tried other settings in between $D_1$, $D_2$ and larger than $D_2$ with consistent results, this is, the difference between {\sc Greedy} and {\sc MGreedy} decreases with the size of the dataset. We do want to show instead how last coalescing time differences change as a function of $d$. For this purpose we generated 250 replicates of $D_1$ and $D_2$ for 6 different values of $d$ and assumed the covariance function parameter as known. Figure~\ref{fg:toy4} shows that both greedy approaches tend to underestimate the last coalescing time, however the mean based algorithm appears to be more accurate and less sensitive to the size of the dataset.

\subsection{Handwritten digits}
The USPS database\footnote{Data available from http://cs.nyu.edu/~roweis/data.html.} contains 9289 grayscale images of digits, each $16\times16$ pixels in size and scaled to fall within the range $[-1,1]$. Here we use subsets of 2500 images, 50 from each digit, randomly selected from the full data set. We apply {\sc MPost2}, {\sc MGreedy} and average-link agglomerative clustering ({\sc HC}) to 25 of such subsets. For the covariance matrix $\bPhi$ we use a Matérn covariance function with parameter $\nu=3/2$ and additive noise defined as follows
\begin{align*}
	g(i,j,\ell_x,\ell_y,\sigma^2) = \left(1+\frac{\sqrt{3}}{\ell_x}d_{x,ij}\right)\left(1+\frac{\sqrt{3}}{\ell_y}d_{y,ij}\right)\exp\left( -\frac{\sqrt{3}}{\ell_x}d_{x,ij}+\frac{\sqrt{3}}{\ell_y}d_{y,ij} \right) + \sigma^2\delta_{ij} \,,
\end{align*}
where $d_{x,ij}$ and $d_{y,ij}$ are distances in the two axes of the image, and we have assumed axis-wise independence \citep{rasmussen06}.
\subsubsection{Performance metrics}
As performance measures we use (i) the \emph{subtree} score defined as $N_{\rm subset}/(n-K)$, where $N_{\rm subset}$ is the number of internal nodes with leaves from the same class, $n$ is the number of observations and $K$ is the number of classes \citep{teh08}, and (ii) the area under the adjusted Rand index (ARI) curve (AUC), which is a similarity measure for pairs of data partitions.  It takes values $\leq 1$ where 1 indicates that the two partitions agree as much as possible given their respective sizes \citep{hubert85a}. 

\begin{figure}[!t]
	\begin{minipage}[c]{0.37\linewidth}
		\centering
			\includegraphics[width=2.2in]{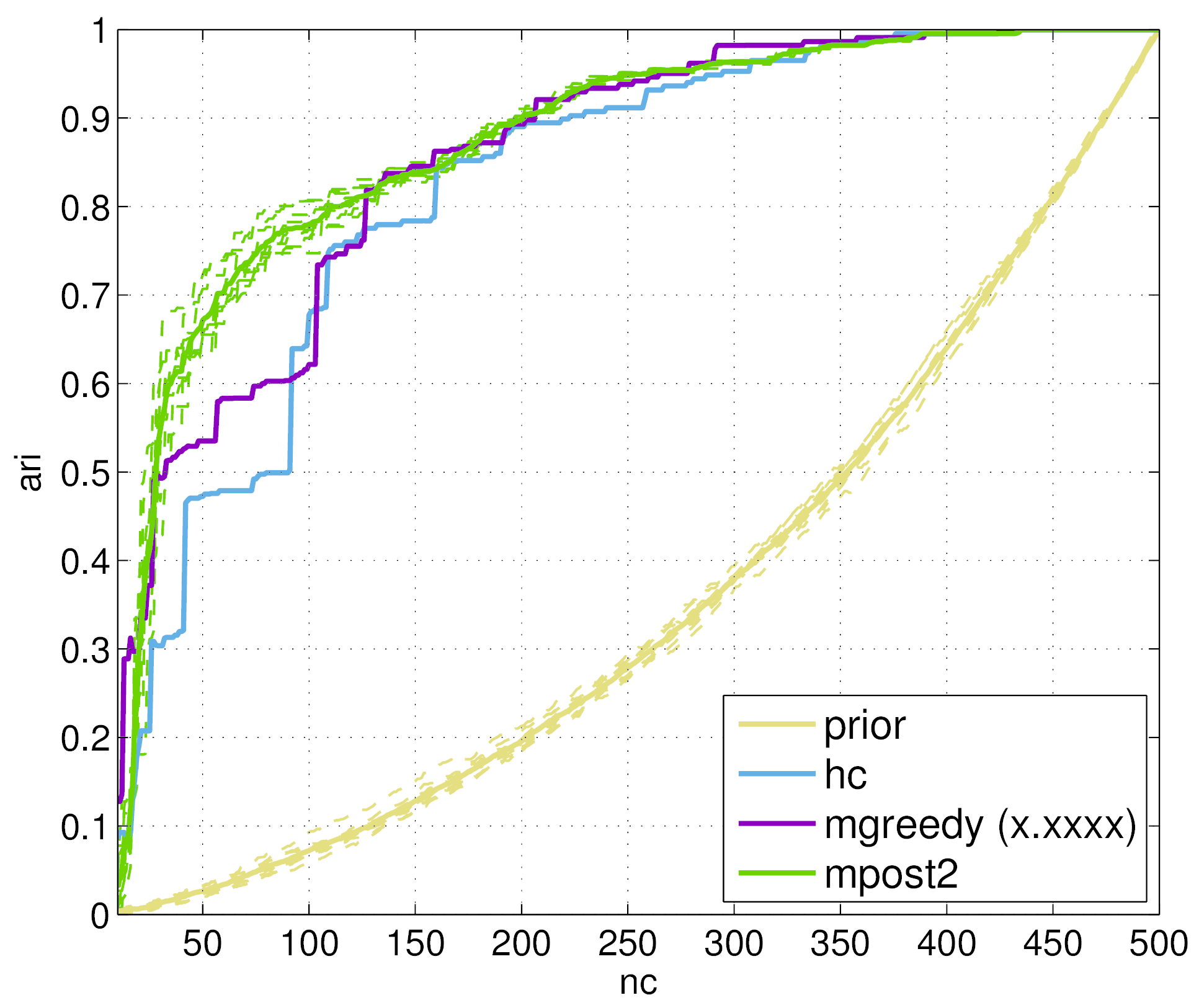}
	\end{minipage}
	\begin{minipage}[c]{0.6\linewidth}
		\centering\small{
		\begin{tabular}{cccc}
			\hline
			& Subtree & AUC & RT \\
			\hline
			{\sc MPost2} & $0.86\pm0.008$ & $ 0.92\pm0.005$ & $1000.30\pm0.17$ \\
			{\sc MGreedy} & $0.85\pm0.002$ & $0.91\pm0.002$ & $26.50\pm0.05$ \\ 
			{\sc HC} & $0.84\pm0.002$ & $0.90\pm0.002$ & $20.00\pm1.00$ \\
			{\sc Ref} & $0.84\pm0.009$ & $0.90\pm0.005$ & $800.30\pm0.17$ \\
			\hline
		\end{tabular}\\
		}
	\end{minipage}
	\caption[USPS digits results]{USPS digits results. (Left) USPS data ARI curves. {\sc Prior} draws structures directly from a $n$-coalescent prior, {\sc Ref} is {\sc MPost2} with diagonal covariance matrix and {\sc HC} is standard hierarchical clustering with average link function and euclidean distance metric. Figures in parenthesis are AUC scores. (Right) Subtree scores, AUC is area under the ARI curve and RT is runtime in minutes. Figures are means and standard deviations across 25 replicates.}
	\label{fg:usps_auc}
\end{figure}

For the ARI metric, we do not compare the true partition and clustering directly.  Given a particular clustering, we label each cluster based on voting by the members of the cluster.  Subsequently, a partition is created by relabeling every observation to match the label of its cluster.  This has the effect of producing perfect accuracy when the partition consists of $n$ singleton sets.   We test all $n$ possible partitions from the clustering model and plot ARI vs number of clusters $N_c$.  This produces a graphical representation that resembles a ROC curve; if performance is perfect, ARI will be 1 for $N_c>K$.  Also, ARI will increase with $N_c$. When $N_c=1$ and $N_c=n$, ARI is always 0 and 1, respectively. Just like in a ROC curve, an algorithm is as good as its ARI's rate of change thus we can asses the overall performance by computing the area under the ARI curve. Figure~\ref{fg:usps_auc} shows curves for a particular data set and the three considered algorithms. We also included results obtained by tree structures drawn from the coalescent prior and {\sc MPost2} with diagonal covariance matrix as baselines.

Table in Figure~\ref{fg:usps_auc} shows average subset scores for the algorithms considered. We performed inference for 20 iterations and 10 particles. No substantial improvement was found by increasing $N_{\rm iter}$ or $M$. {\sc Greedy} produced the same results as {\sc MGreedy} and {\sc SMC1} did not performed better than {\sc MPost1/2} but it took approximately 10 times longer to run (results not shown). We see that coalescent based algorithms perform better than standard hierarchical cluster in terms of AUC. {\sc MPost1} and {\sc MPost2} are best in subtree scores and AUC, respectively.
\subsection{Motion capture data}
\begin{figure}[!t]
	\centering
			\includegraphics[width=2.5in]{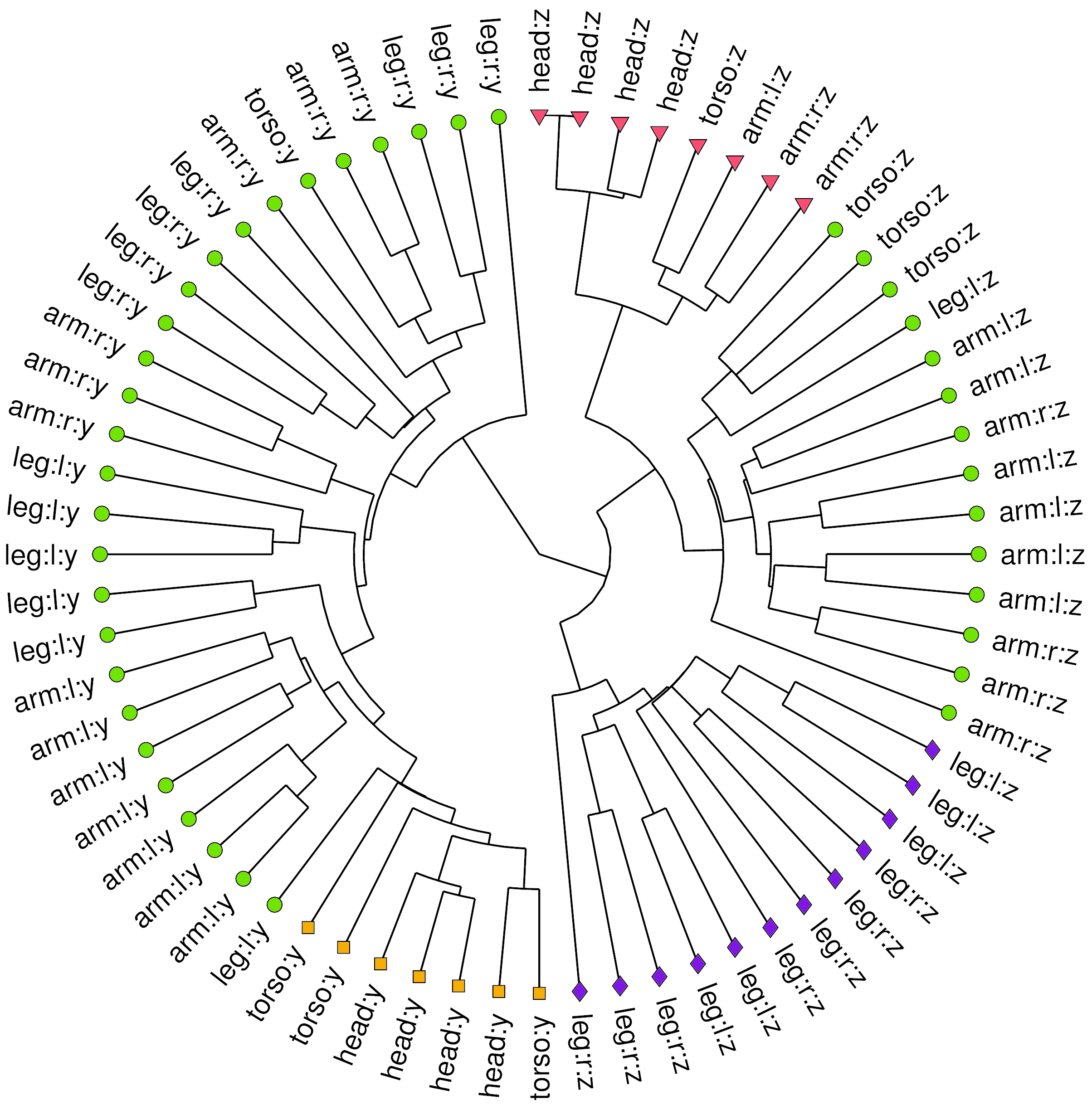} \hspace{2mm}
			\includegraphics[width=2.5in]{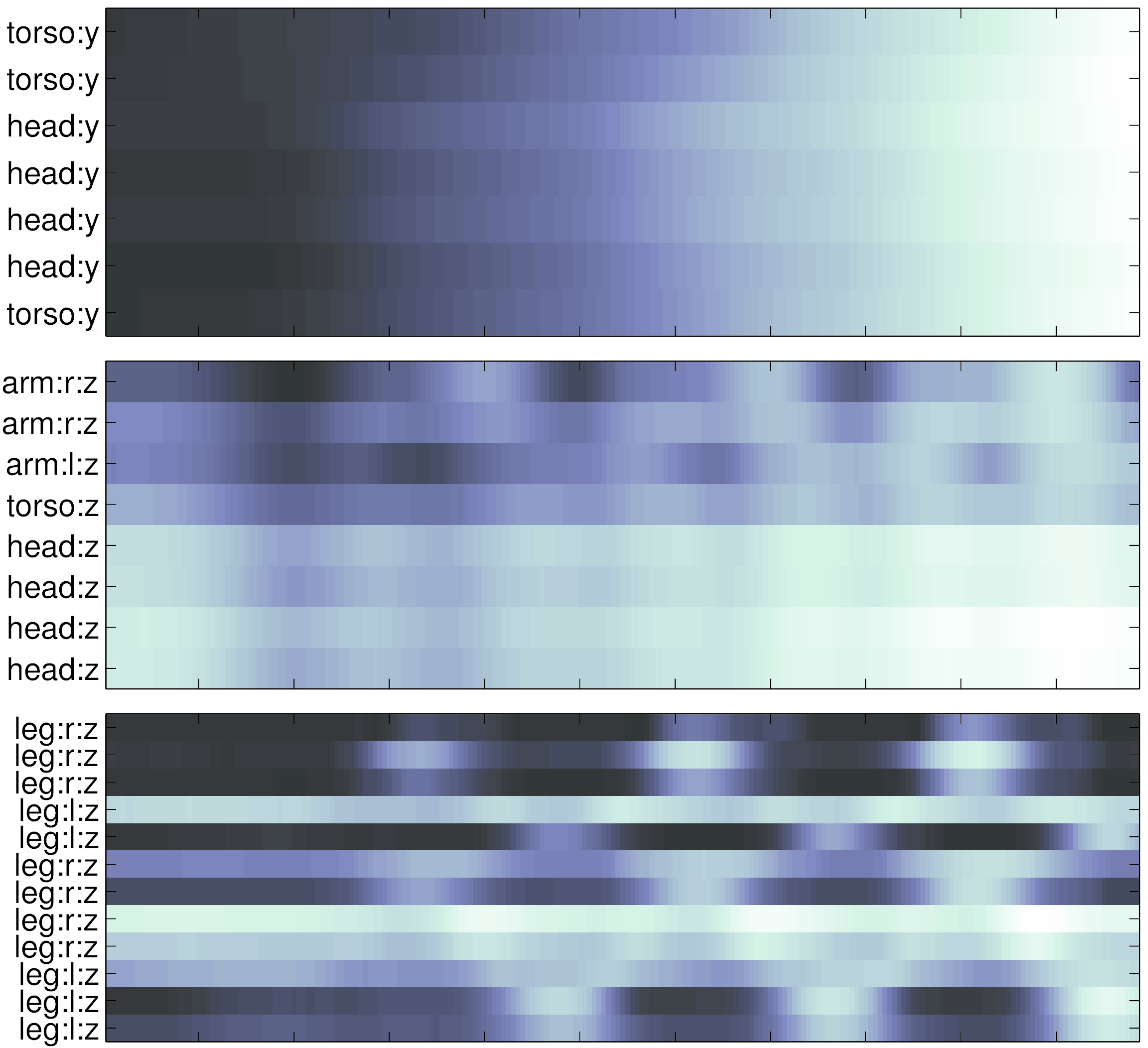}
	\caption[MOCAP data results]{MOCAP data results. (Left) Resulting subtree from the particle with maximum weight (0.0784) at iteration 50. (Right) Data corresponding to three subtrees of (Left) marked with squares, triangles and diamonds, respectively.}
	\label{fg:mocap_tree}
\end{figure}

We apply {\sc MPost2} to learn hierarchical structures in motion capture data (MOCAP). The data set consists of 102 time series of length 217 corresponding to the coordinates of a set of 34 three dimensional markers placed on a person breaking into run\footnote{Data available from http://accad.osu.edu/research/mocap/mocap\_data.htm.}. For the covariance matrix $\bPhi$, we used the squared exponential function in equation~\eqref{eq:sqexp_noise}. Results are obtained after running 50 iterations of {\sc MPost2} with 50 particles. It took approximately 5 minutes to complete the run. Left panel in Figure~\ref{fg:mocap_tree} shows two subtrees containing data from all markers in the $Y$ and $Z$ axes. 

In order to facilitate visualization, we relabeled the original markers to one of the following: head (Head), torso (Torso), right leg (Leg:R), left leg (Leg:L), right arm (Arm:R) and left arm (Arm:L). The subtrees from Figure~\ref{fg:mocap_tree} are obtained from the particle with maximum weight (0.129) at the final iteration of the run with effective sample size 24.108. We also examined the trees for the remaining particles and noted no substantial structural differences with respect to Figure~\ref{fg:mocap_tree}. 

The resulting tree has interesting features: (i) Sensors from the $Y$ and $Z$ coordinate axes form two separate clusters. (ii) Leg markers have in general larger merging times than the others, whereas the opposite is true for head markers. (iii) The obtained tree agrees with the structure of the human body reasonably well; for instance in the middle-right panel of Figure~\ref{fg:mocap_tree} we see a heat map with 9 markers, 4 of them from the head, 1 from the torso (C7, base of the neck) and 4 from the arms (shoulders and upper arms). The two arm sensors close to the torso correspond to the shoulders while the other two---with larger merging times, are located in the upper arms. 

The obtained structure is fairly robust to changes in the number of iterations and particles. {\sc MPost1}, {\sc mGreedy}, {\sc SMC1}, and {\sc Post-Post} produce structurally similar trees to the one shown in Figure~\ref{fg:mocap_tree}, however with different running times. In particular, they took 7, 1, 12 and 75 minutes, respectively.

\bibliography{mlbib}
%
\end{document}